\RequirePackage{fix-cm}
\documentclass[smallextended]{svjour3}       
\usepackage{graphicx}
\usepackage{listings}
\usepackage{color}
\usepackage[utf8]{inputenc}
\usepackage{amsmath}
\usepackage{tikz}
\usepackage{esvect}
\usepackage{amsfonts}
\usepackage{bm}
\usepackage[font={small,it}]{caption}
\usepackage{subcaption}
\usepackage{booktabs}
\usepackage[
backend=biber,
]{biblatex}

\usepackage{float}%
\floatstyle{plaintop}%
\restylefloat{table}

\usepackage{multirow}

\usepackage{diagbox}

\usepackage{xcolor,colortbl}
\usepackage{transparent}

\usepackage{enumitem}
\setlist{topsep=1.5em, itemsep=1.5em}

\definecolor{Gray}{gray}{0.9}
\definecolor{LightGray}{gray}{0.95}
\newcolumntype{a}{>{\columncolor{Gray}}c}
\newcolumntype{b}{>{\columncolor{LightGray}}c}

\newcommand{\irow}[1]{
  \begin{smallmatrix}(#1)\end{smallmatrix}%
}
\definecolor{dkgreen}{rgb}{0,0.6,0}
\definecolor{gray}{rgb}{0.5,0.5,0.5}
\definecolor{mauve}{rgb}{0.58,0,0.82}
\bibliography{sbc-template.bib}
\begin{document}

\title{A Model-Based Approach To Assess Epidemic Risk
}
%



\author{Hugo Dolan          \and
        Riccardo Rastelli 
}


\institute{Hugo Dolan \at
              University College Dublin \\
              \email{hugo.dolan@ucdconnect.ie}           
           \and
           Riccardo Rastelli \at
              University College Dublin \\
              \email{riccardo.rastelli@ucd.ie} 
}

\date{Received: date / Accepted: date}

\maketitle

\begin{abstract}
We study how international flights can facilitate the spread of an epidemic to a worldwide scale. 
We combine an infrastructure network of flight connections with a population density dataset to derive the mobility network, and then we define an epidemic framework to model the spread of the disease. 
Our approach combines a compartmental SEIRS model with a graph diffusion model to capture the clusteredness of the distribution of the population.
The resulting model is characterised by the dynamics of a metapopulation SEIRS, with amplification or reduction of the infection rate which is determined also by the mobility of individuals.
We use simulations to characterise and study a variety of realistic scenarios that resemble the recent spread of COVID-19.
Crucially, we define a formal framework that can be used to design epidemic mitigation strategies: we propose an optimisation approach based on genetic algorithms that can be used to identify an optimal airport closure strategy, and that can be employed to aid decision making for the mitigation of the epidemic, in a timely manner.
\keywords{COVID-19 \and SEIRS compartmental model \and Genetic Algorithm \and Network Analysis \and Human Mobility}
\end{abstract}

\section{Introduction}
In recent years, the extensive development of the transportation infrastructure has radically changed how connected our world is. 
International flights allow individuals to travel around the globe in just a few hours or days. 
This has important negative implications on the spread of diseases, whereby epidemics can reach a worldwide scale before effective responses are set in place. 
The recent COVID-19 outbreak \cite{shereen2020covid} has clearly raised and emphasised this problem. 
As a response to the emergency, many countries have taken drastic measures to contain and slow down the spread of the virus by imposing lockdowns and airport closures \cite{chinazzi2020effect}. While these measures have been successful in confining the epidemic, the immediate and chaotic response has blurred the actual role played by the topology of the infrastructure network on the spread of the virus. 

The main goal of this paper is to create a model-based framework that can inform decision making regarding airport closures as a means to slowing down a potential epidemic without causing excessive economic damage. 
In particular, we introduce a new framework to study networks of international flights as potential vehicles for the spread of diseases. 

In this paper, we first propose an in-depth analysis of the Open Flights network dataset \cite{openflights}, which describes a large number of flight connections between more than $3{,}000$ airports. 
We calculate a number of descriptive statistics from the data, in order to study the underlying topology of this infrastructure network, and essentially to understand how individuals can move between distant locations.
We use various centrality measures to identify key airports, and we test the resilience of the network when these key airports are removed.

Then, we fit a Stochastic BlockModel (SBM) to partition the airports into homogeneous groups.
The SBM, originally introduced and studied by \cite{wang1987stochastic}, is a fundamental model and tool for statistical network analysis, since it can highlight groups of nodes that exhibit similar connectivity patterns. 
Inference for the SBM can be performed using both classical and Bayesian approaches \cite{nowicki2001estimation}. 
One fundamental aspect of this model is that it can be interpreted as a finite mixture model for networks \cite{daudin2008mixture}, and thus it borrows many concepts and tools from this related research area.
In this context, a useful by-product of the SBM framework is that it allows us to compare the connectivity and clustering of the airports with their actual geographical location.

After the exploratory data analysis, we use the infrastructure network to create a model for the simulation of epidemics. 
An essential aspect of this task is the development of a statistical network model that can combine these flight routes data with the geographical distribution of the population. 
Our aim is to give a model-based quantification of the epidemic risk which is amplified by the travelling of individuals, and to possibly identify effective interventions that can mitigate this risk. 
\cite{chinazzi2020effect} propose an approach similar to ours, in that they combine the infrastructure network with the gridded population data to study the effects of the airport closure interventions that were actioned at the beginning of this year.
\cite{chinazzi2020effect} use a tool called GLEAM \cite{balcan2009multiscale,balcan2010modeling} which can combine data from different sources to predict the behaviour of the epidemic using an individual-based compartmental SEIRS model.
While they focus on the effects following the actual airport closures, in this paper we aim at defining a framework to take new decisions that can lead to optimal airport closures, or potential future airport re-openings.

Our approach relies on epidemic compartmental models \cite{hethcote2000mathematics}, and in particular on a SEIRS model.
This framework postulates that the population is divided in $4$ ordered compartments (susceptible, exposed, infectious, recovered) and that different rates determine the flow of individuals from one compartment into the next, and eventually back into the first compartment.
This family of models has been largely employed in various research fields to model the evolution of epidemics, and it has been also successfully used within the context of COVID-19 \cite{hou2020effectiveness, chinazzi2020effect, peng2020epidemic, radulescu2020management, lopez2020modified}.

One fundamental aspect of our epidemic model is that, similarly to \cite{chinazzi2020effect, tizzoni2014use, apolloni2014metapopulation}, we consider a metapopulation where each subpopulation is centred at an airport location, and whereby the local epidemic is determined by a SEIRS model. 
We use a graph diffusion process to describe the flows between the various subpopulations, which in turn affect the local dynamics.
Not only this allows us to observe the epidemic both locally (for each subpopulation) and globally, but it also allows us to appreciate the spatio-temporal progression of the virus.

We calibrate the parameters of our model so that our predictions are similar to the recent spread of COVID-19.
Our method is flexible and it can be used for a diverse range of epidemic parameters, hence encompassing other relevant diseases, or COVID-19 variants of interest.
We do not claim that our results are specific to the COVID-19 epidemic nor that they should be used within this context; rather, we provide a general methodology that could be employed in any epidemic setting, and, for example purposes, we recreate realistic situations that resemble the recent COVID-19 epidemic.
In fact, we test the sensitivity of our model by running a number of simulations that encompass a variety of possible real epidemic scenarios. 

The fundamental contribution of this work regards the study of optimal epidemic mitigation strategies. 
Once we possess a model which is calibrated to a realistic setting, we explore an optimisation approach to identify what could be the optimal airport closure strategies that should be implemented. 
We use the predictions from our epidemic model to construct an objective function that takes into account measures for the spread of the disease as well as economic losses.
We perform the optimisation using Genetic Algorithms (GAs) \cite{spall2005introduction}.
GAs are heuristic stochastic optimisation algorithms that explore new candidate solutions by selecting and transforming a set of current solutions using some basic principles of evolution and natural selection.
In our context, GAs are especially convenient since the problem that we address is a combinatorial one, where we want to find the optimal subset of airports that should be closed to minimise our objective function.

The software that we have used in this paper is maintained by the authors and is available from the GitHub repository: \cite{github}.

\section{Network Topology} \label{sec:NetworkTopology}
\begin{figure}[htpb]
\centering
\includegraphics[width=1\textwidth]{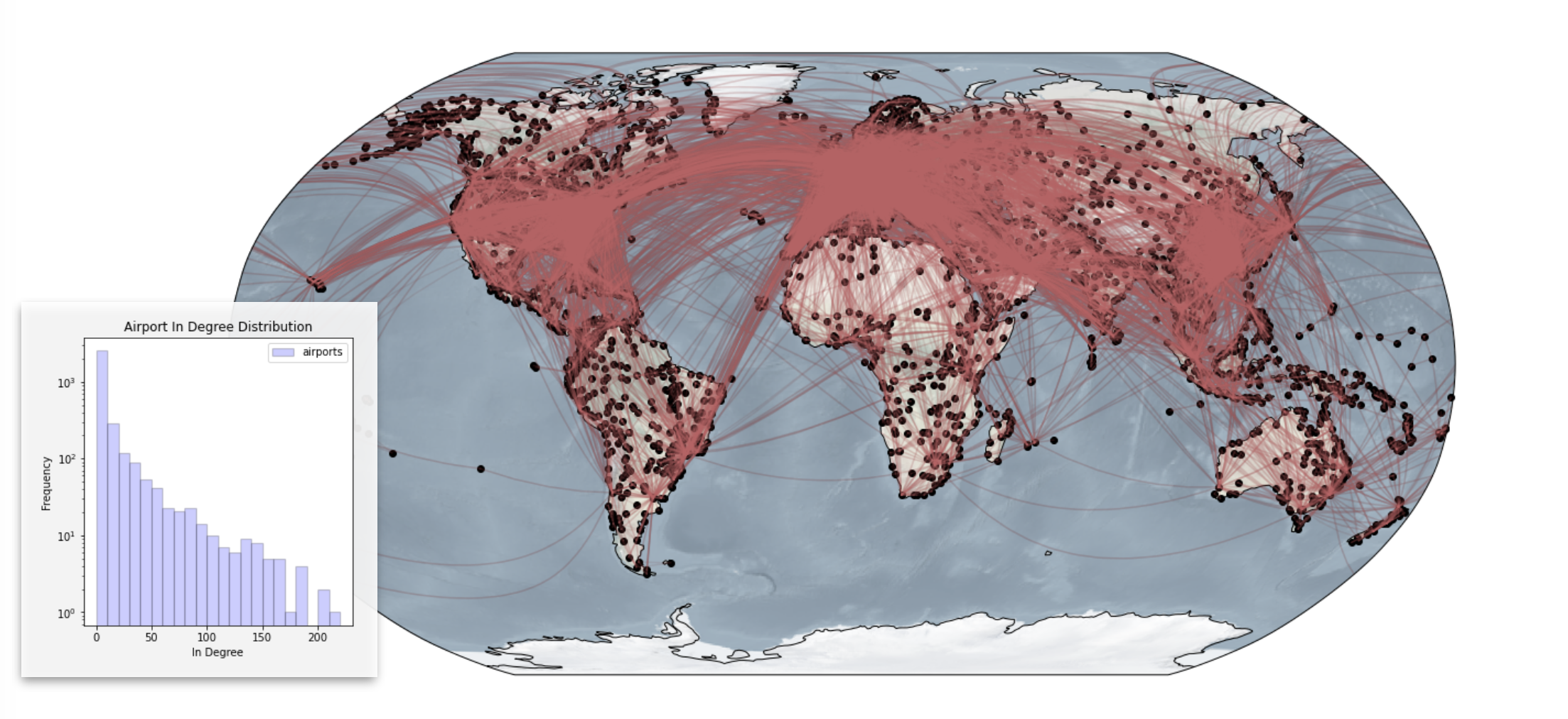}
\caption{Open Flights Network Visualisation and its in-degree distribution.}
\label{fig:NetworkDiagram}
\end{figure}

In this section we propose an exploratory data analysis and basic statistical modelling of the Open Flights dataset \cite{openflights}.
The Open Flights dataset contains information on $3{,}425$ airports globally, including a database of $37{,}594$ commercial routes between these airports collected in 2014. 
The dataset is transformed into an adjacency matrix with nodes representing airports and directed edges representing whether there exists a direct route between any two airports. 
In Figure \ref{fig:NetworkDiagram} we present the network visually, and, on initial inspection, it is clear that the network exhibits extremely high degree of connectivity, with the plot of degree distribution indicating that over $20\%$ exhibit a degree greater than $10$. 

Identifying airports which are important to the overall connectivity of the network is crucial in gaining a better understanding of the network’s topology. 
We consider several metrics to measure the importance of nodes. 
These include the Page-Rank centrality, betweenness centrality, coreness ranks, as well as the in-degrees and out-degrees of nodes \cite{mark:10}.
We present a table of the $20$ most important airports according to Page-Rank in Table \ref{tab:NetworkSummary}.  
\begin{table}
\centering
\caption{Summary Statistics for the top $20$ airports, sorted according to Page-Rank centrality values. Additional node statistics and centralities are shown in the other columns, including in-degree ($\textbf{deg(+)}$) and out-degree ($\textbf{deg(-)}$).}
\label{tab:NetworkSummary}
\includegraphics[width=0.9\textwidth]{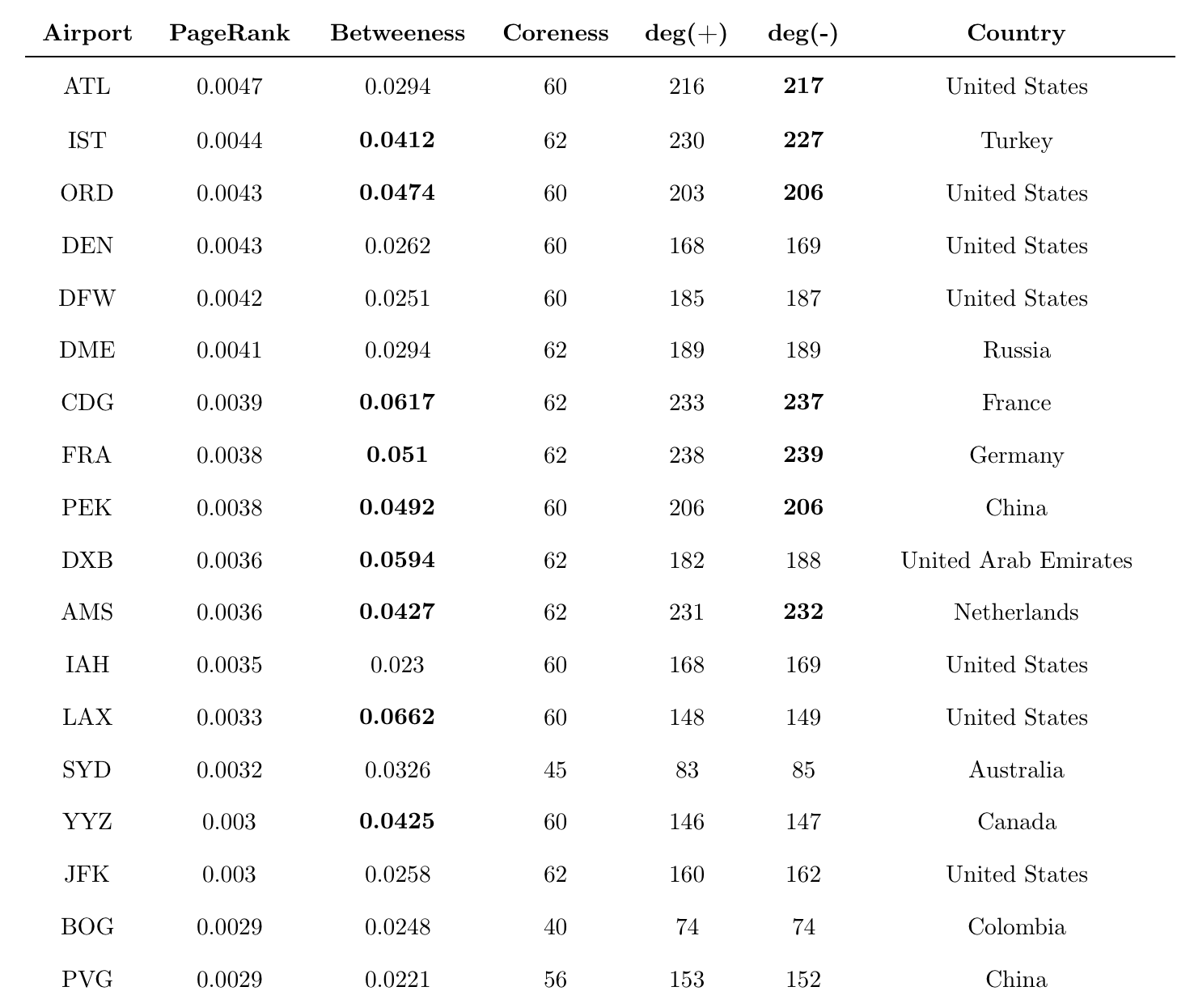}
\end{table}
Airports with high Page-Rank are clearly major international destinations, and they form an extremely well connected subnetwork with a coreness of over $60$, meaning that every airport in this sub-network has $60$ or more connections, or, equivalently, since the graph is approximately symmetric (the majority of air routes run return flights), we can say that every airport in this subnetwork has an out-degree close to, or exceeding, $30$. 
It is interesting to note that airports with high betweenness centrality (Charles De Gaulle, Dubai, Beijing, Amsterdam, Los Angeles, Toronto, Frankfurt) are also major connecting flight hubs, exhibiting out-degrees of over $200$. 

\begin{figure}[htpb]
	\centering
	\begin{subfigure}{0.9\textwidth}
		\centering
		\includegraphics[width=0.9\textwidth]{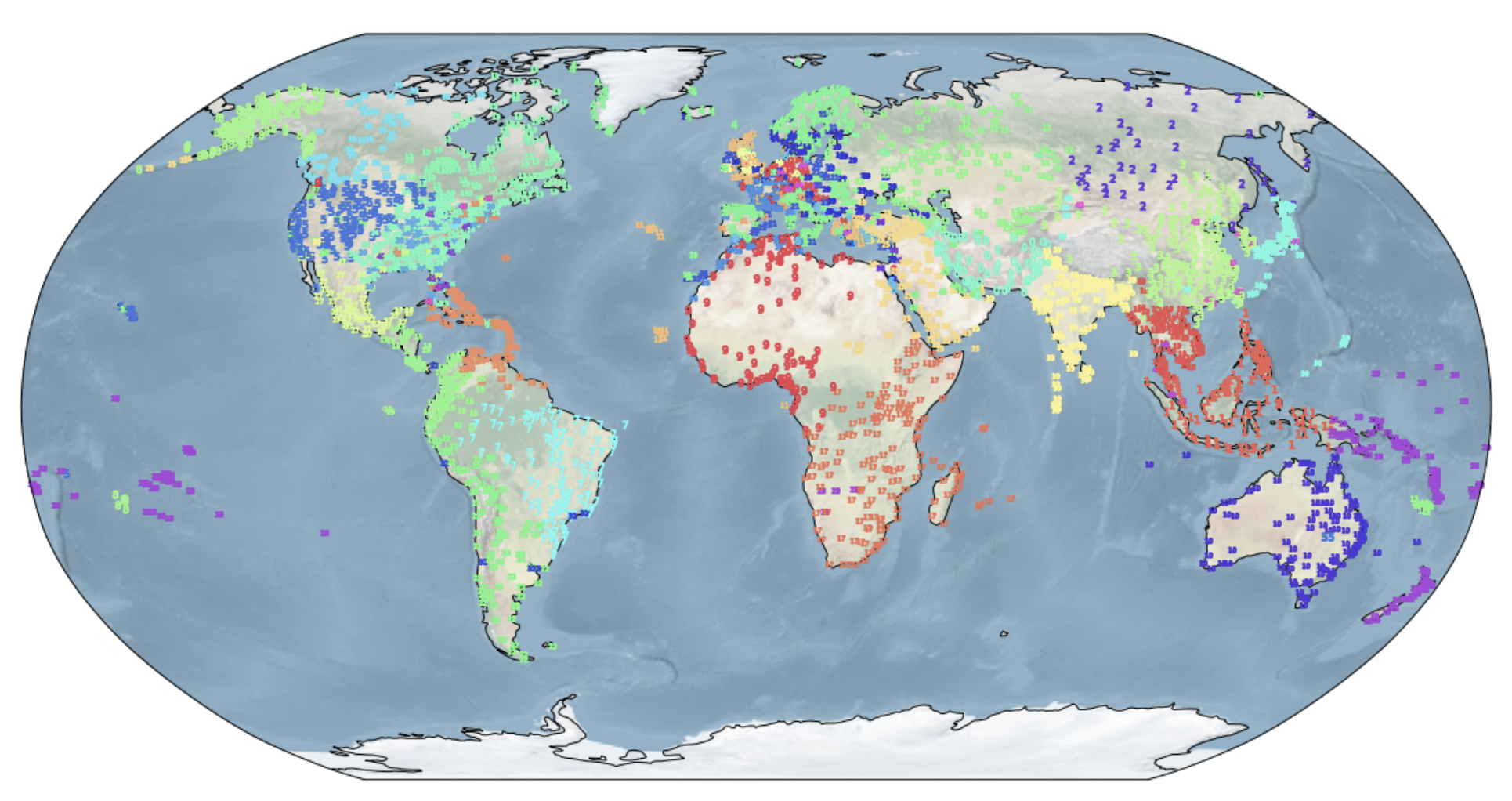}
		\caption{Map of the SBM results, with colours indicating the cluster memberships.}
		\label{fig:BlockStochastic}
	\end{subfigure} \\
	\begin{subfigure}{0.9\textwidth}
		\centering
		\includegraphics[width=0.8\textwidth]{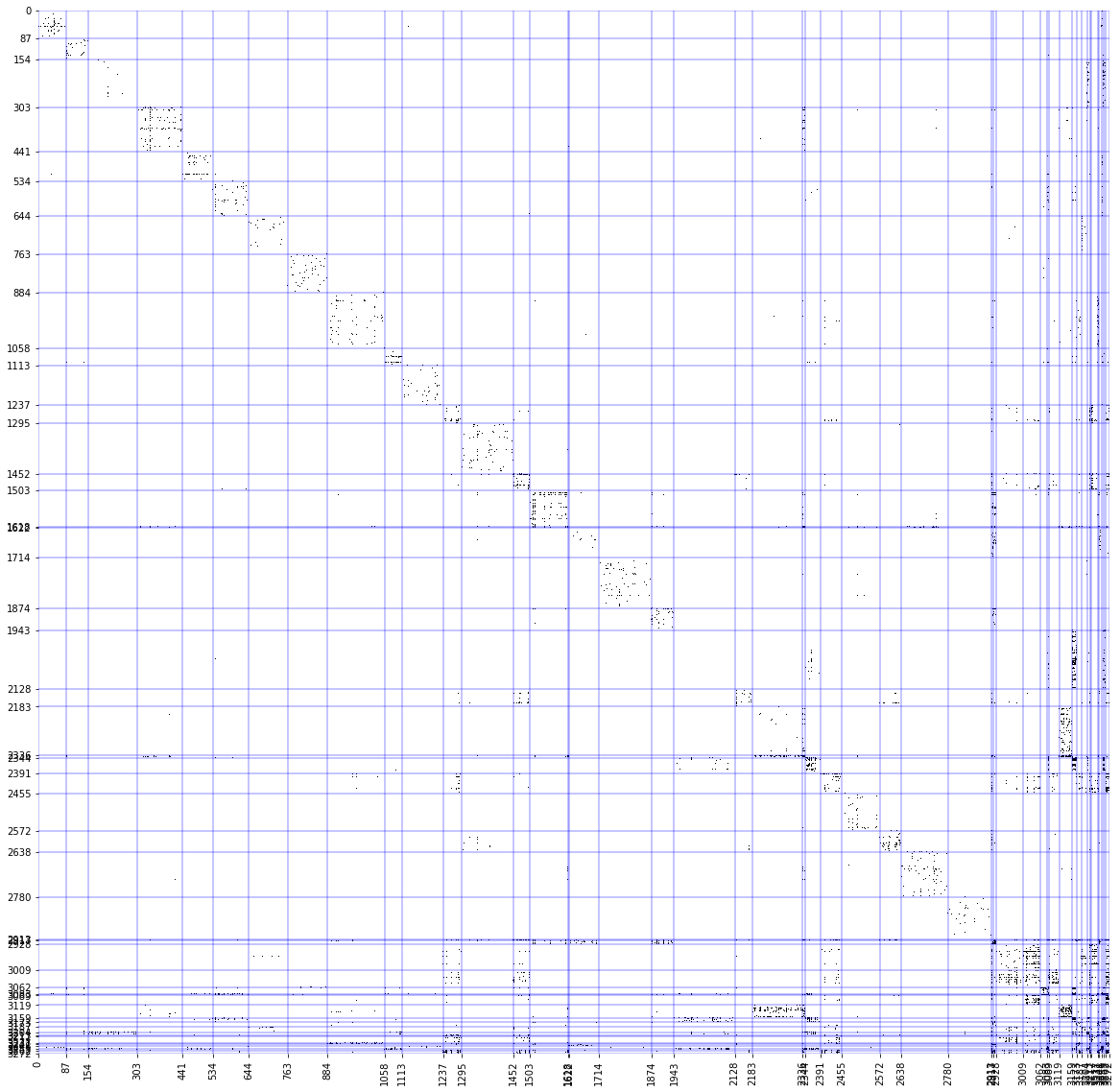}
		\caption{Block matrix visualisation of the adjacency matrix, where the black dots indicate routes between airports and the blue boxes indicate the clusters.}
		\label{fig:DotPlot}
	\end{subfigure}
	\caption{Fitted SBM results.}
\end{figure}

We identify homogeneous subgroups of airports within the network by employing a SBM framework: we use a Python implementation of an efficient Markov chain Monte Carlo method, which is suitable for inferring SBMs in large networks, as described by \cite{Peixoto_2014}. 
The optimal SBM partition and the corresponding block matrix are shown in Figures \ref{fig:BlockStochastic} and \ref{fig:DotPlot}, respectively.
It is clear that the communities found are very strongly associated with the geographical location of the airports, and with their region or province. 
This is quite surprising as this information is not encoded explicitly in the data provided to the algorithm. 
This would strongly suggest a high degree of connectivity of airports not only globally but also within regions or geographical areas. 
We also note from the block matrix in Figure \ref{fig:DotPlot} that the majority of connections are within relatively large communities representing the geographic clustering observed in Figure \ref{fig:BlockStochastic}, but also towards the lower right corner there is significant disassortative behaviour, likely these nodes are large international hubs such as the small community of London, Frankfurt, Amsterdam, Charles De Gaulle which share connections to many cities across the world. 

\begin{figure}[htpb]
\centering
\includegraphics[width=0.9\textwidth]{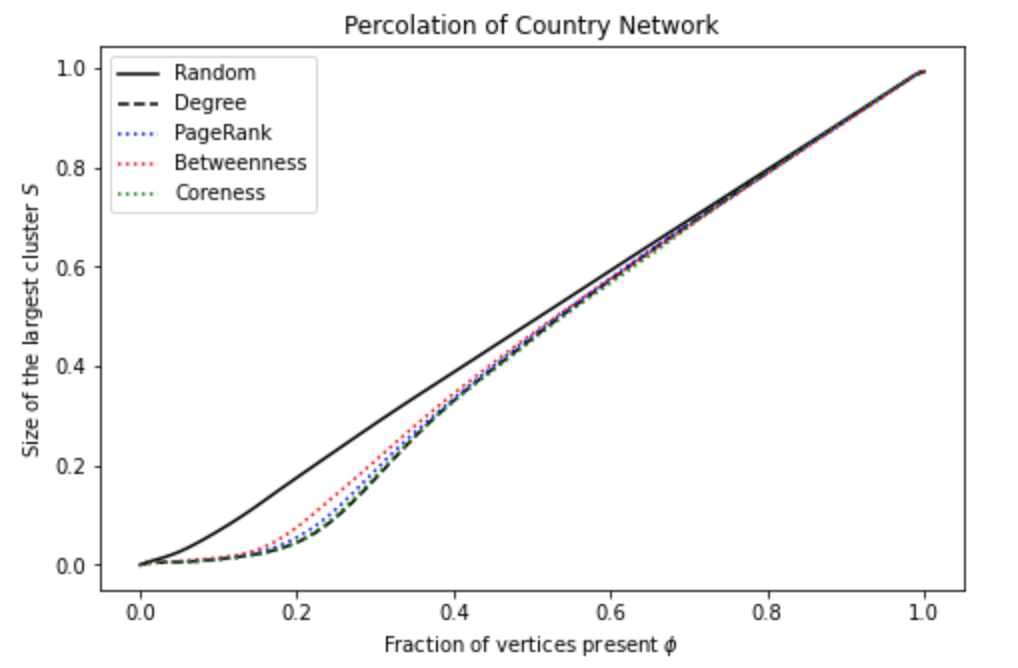}
\caption{Percolation on the Open Flights dataset via a variety of ranking criteria (results are averaged over $100$ independent simulations).}
\label{fig:Percolation}
\end{figure}

We continue our exploratory analysis by studying the percolation properties of the network \cite{mark:10}.
We percolate the network by sequentially removing the nodes (one at a time) and their connections, and observing how the connectivity of the graph changes.
We remove the nodes following both a random order, and following a decreasing order of the out-degree and other centrality measures.
The results are shown in Figure \ref{fig:Percolation}. 
The network is highly resilient to random attacks, since the removal of almost all nodes is required in order to disrupt network connectivity. 
However, we note that the network is moderately more vulnerable to targeted (degree-based) attacks, yet this would still require more than half of all airports to be removed for the single giant component to disappear. 
Similarly, the network is moderately vulnerable to targeted attacks according to other ranking factors (Page-Rank, betweenness, coreness).
We note that these procedures are also averaged over many trials to account for the removal of vertices of equal rankings in different orders, however, we find these results very quite similar to percolation by degree.

In conclusion, the Open Flights network summary statistics show that airports which are large regional destinations, or hubs for connecting flights, tend to have high importance to network connectivity. 
Furthermore, it is observed that some nodes in the network are extremely well connected, both at regional and global level, with significant geographical community structure. 
The network is also highly resilient to random or deterministic attacks. 

In the context of epidemics, these initial findings provide a very solid evidence that the flight connections can sadly be a very efficient vehicle to facilitate the spread of diseases, and, more importantly, that substantial network ``damage'' (e.g. airport closures) is required to ensure that an epidemic does not spread to a worldwide scale.
This evidence motivates our work, in that we aim at finding optimal mitigation strategies that can reduce the pace of the epidemic to a much smaller scale, without causing excessive disruption to economies and to this particular infrastructure network.

\section{Model Specification} \label{sec:ModelSpecification}

\subsection{Theoretical Underpinnings}  \label{sec:Theory}
Before we develop the main model of this paper, we must first introduce two existing models which can be found in the domains of applied mathematics and epidemiology \cite{mark:10}. 
Firstly, we specify the graph diffusion model, which describes the flow of a fluid across a network, driven by pressure differences between adjacent nodes. 
This can be expressed as a vector of differential equations denoting changes of fluid volumes at each node and time step:
$$\frac{d\boldsymbol{\psi}}{dt} = c(\textbf{A} - \textbf{D})\boldsymbol{\psi}$$
We use the notation \boldmath{$\psi$} to represent the vector of fluid volumes at every node, \boldmath{$A$} to denote the adjacency matrix of the network, \boldmath{$D$} to denote a diagonal matrix of containing the degrees of every node, and $c$ is called the diffusion constant. 
A full derivation of this can be found in \cite{mark:10}. 

\begin{figure}[htpb]
\centering
\includegraphics[width=0.5\textwidth]{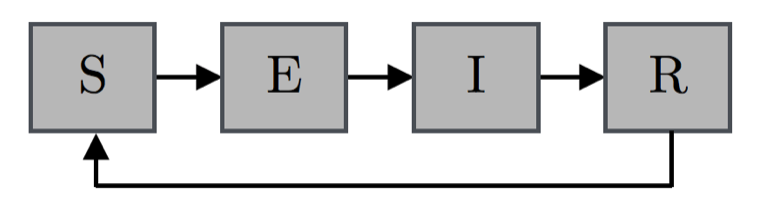}
\caption{SEIRS compartmental flow diagram.
}
\label{fig:DiffEquations}
\end{figure}

Additionally, we introduce the SEIRS compartmental epidemiology model. Each letter of the model name denotes a compartment of the system (Susceptible ($S$), Exposed ($E$), Infectious ($I$) and Recovered ($R$)), in which some number of individuals from the total population ($M$) reside. Figure \ref{fig:DiffEquations} illustrates the direction of progression from state to state, whilst \eqref{eq:seirs_1} indicates the exact rates at which the population in each compartment changes. 
\begin{equation}\label{eq:seirs_1}
\begin{split}
 \frac{dS}{dt} &= \delta R - \frac{S\beta I}{M} \\
 \frac{dE}{dt} &= \frac{S\beta I}{M} - \epsilon E \\
 \frac{dI}{dt} &= \epsilon E - \gamma I \\
 \frac{dR}{dt} &= \gamma I - \delta R \\
\end{split}
\end{equation}
The epidemic parameters $\beta$, $\epsilon$, $\gamma$ and $\delta$ are non-negative scalars, and can be estimated or calibrated to match the characteristics of some observed epidemic. Once the system's initial condition and the epidemic parameters are specified, we can find a numerical approximation to the solution of this system, hence obtaining the number of individuals in each compartment, at each time $t>0$.

\subsection{Model Definition} \label{sec:ModelDef}
In order to model the transmission of disease through the international flights network, we opt to use the SEIRS model combined with a graph diffusion model as described in the previous section. We will refer to the airports' adjacency matrix as $A$ and denote airport nodes as \boldmath$v_j$, for $j = 1,\ldots,N.$ The total number of nodes in the network is \boldmath$N$ and the associated population at each node is \boldmath$M_j$. Let us define the local epidemic state vector as $\theta_j(t) = \irow{S_j & E_j&I_j&R_j}^\top$, which represent the compartments of the SEIRS model for airport population at any given time. A condition of the SEIRS model constrains the total population of all compartments to equal the total population, at each time $t$:
$$ S_j(t) + E_j(t) + I_j(t) + R_j(t) = M_j(t) $$
We assume that the local population is fully mixed (i.e. everyone has equal chance of being infected), as this is a standard assumption of compartmental epidemic models, however we assume this only to be true at the individual airport level, and not for the entire global system. Additionally we introduce \boldmath$\alpha_j$ as the proportion of the population which can travel, and \boldmath$c$ as the probability that an individual departs from an airport on any given day. The proportions allow us to define an additional variable \boldmath$\psi_j(t) = \alpha_j \theta_j(t)$ which corresponds to the mobile epidemic state. 

We can now define, at a high level, our simulation procedure, which we use to generate the SEIRS data conditionally on a specific set of epidemic parameters: 
\begin{enumerate}
  \item At each time-step we must first initiate community spread of the disease through each airport's local population. This is represented by the compartments of its SEIRS model denoted by the local epidemic state vector $\theta_j$. Using a fourth-order Runge-Kutta approximation technique we update each compartment's value in accordance with the dynamics of the SEIRS model. The new state of the compartments after community spread is denoted $\theta^*_j$ as opposed to the previous state $\theta_j$, for all $j$s.
  
  \item Now take this new local epidemic state $\theta^*$ and split it into two types of population members: a base population denoted $\theta_B$, who are permanent residents to the local area (i.e. the \textbf{locals}); and a second group $\theta_T$, identifying the transient population who are temporary \textbf{visitors}, who have arrived at the airport on business or holidays and will return home after a short period. This differentiation is necessary to ensure that the local populations remain stable over time (we assume no permanent migration in our model).
  
  \item Next, we compute the proportions of $\theta_B$ and $\theta_T$ who can fly (in each compartment), these are $\psi_+ = \alpha_+ \theta_B$ and $\psi_- = \alpha_- \theta_T$, describing the number of outbound passengers and returning travellers, respectively.
  
  \item Use a diffusion model to compute the changes of $\theta_B$ and $\theta_T$ at each airport. The diffusion model is tuned to take into account several factors including: the differences between outbound and returning passengers at every connected airport, the relative importance of the airports in the network, and the airport’s number of connections. These guarantee that the diffusion model can affect the SEIRS statuses without changing the distribution of the population across locations.
  
  \item Recombine the updated values of $\theta_B$ and $\theta_T$ into the aggregate populations $\theta$ and repeat the whole procedure for the number of iterations (``days'') required.
\end{enumerate}

\begin{figure}[htpb]
\centering
\includegraphics[width=0.9\textwidth]{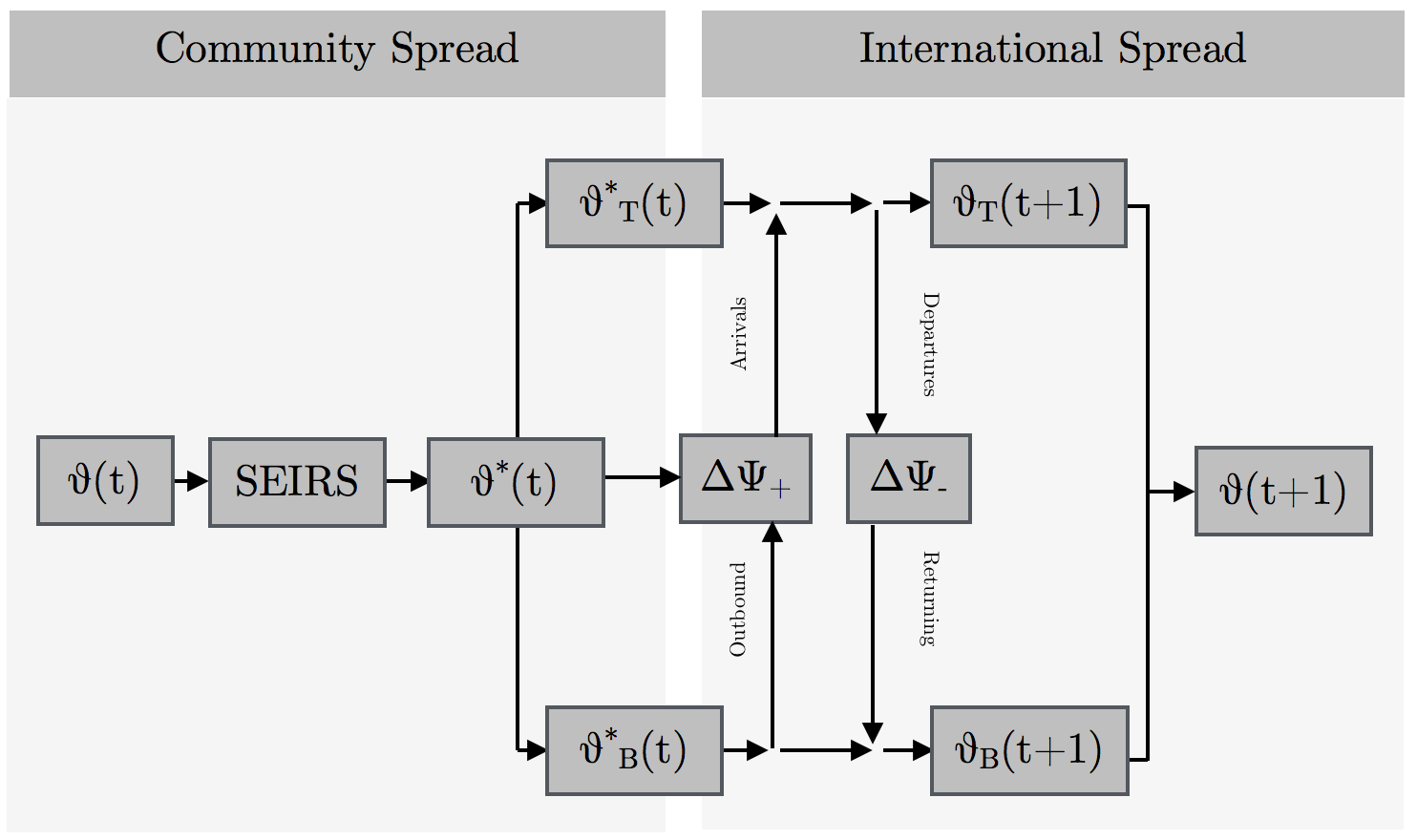}
\caption{Diagram of the epidemic diffusion model steps.}
\label{fig:ModelDiagram}
\end{figure}

For completeness we include both a diagrammatic form of the algorithm (Figure \ref{fig:ModelDiagram}) as well as a the full algorithm (Table \ref{tab:Psuedocode}) which reflects the high level overview above. 
\begin{table}[htpb]
\scriptsize
	\centering
	\caption{Simulation pseudocode, where with $\Sigma$ we indicate a matrix representation of the SEIRS process.}
	\label{tab:Psuedocode}
\bgroup
\def\arraystretch{1.5}
\begin{tabular}{llll}
\toprule
\rowcolor{white}
\multicolumn{4}{l}{\textbf{Algorithm pseudocode for $T$ days}} \\
\midrule
       \multicolumn{3}{l}{\textbf{for} $t = 1, \dots, T:$} & \\
   & & &  \\
\rowcolor{LightGray}
   & & $\theta^*(t-1) = \theta(t-1) + \Sigma\ \theta(t-1)$ & Community spread\\ 
\rowcolor{LightGray}
   & & &  \\
   & & $\theta^*(t-1) = \theta^*_B(t-1) + \theta^*_T(t-1)$ & Preparation for diffusion\\
   & & $\psi_+^*(t-1) = \theta^*_B(t-1)\ \alpha_+$ &\\
   & & $\psi_-^*(t-1) = \theta^*_T(t-1)\ \alpha_-$ & \\ 
   & & & \\
\rowcolor{LightGray}
   & & $\Delta\psi_+(t) = -c_+\psi^*_+(t-1) \left( I - C^\top \right)$ & International spread\\
\rowcolor{LightGray}
   & & $\Delta\psi_-(t) = -c_-\psi^*_-(t-1) \left( I - C^\top \right)$ & \\
\rowcolor{LightGray}
   & & $\theta_B(t) = \theta^*_B(t-1) + Min\left( \Delta\psi_+(t), 0 \right) + Max\left( \Delta\psi_-(t), 0 \right)$ & \\
\rowcolor{LightGray}
   & & $\theta_T(t) = \theta^*_T(t-1) + Max\left( \Delta\psi_+(t), 0 \right) + Min\left( \Delta\psi_-(t), 0 \right)$ & \\
\rowcolor{LightGray}
   & & &  \\
\bottomrule
\end{tabular}
\egroup
\end{table}
A full derivation is provided in Appendix \ref{Algorithm_Derivation}.

\subsection{Modelling assumptions}
In this section we provide further information on the parameters of our model, on their interpretation, and on how we have used different data sources to identify a range of plausible values for these parameters.
We specifically focus on the dataset that have been used and where they come from, and how we have transformed these datasets to obtain the key elements for our study.

Our framework requires information on airports, routes, demographics, wealth, which is difficult to obtain and use. 
A summary of the model’s parameters is provided in Table \ref{tab:KeyParameters}, for convenience. 
\begin{table}[htbp]
\centering
\caption{Model parameters
[*] necessary conditions for the escalation of the epidemic.}
\label{tab:KeyParameters}
\includegraphics[width=1\textwidth]{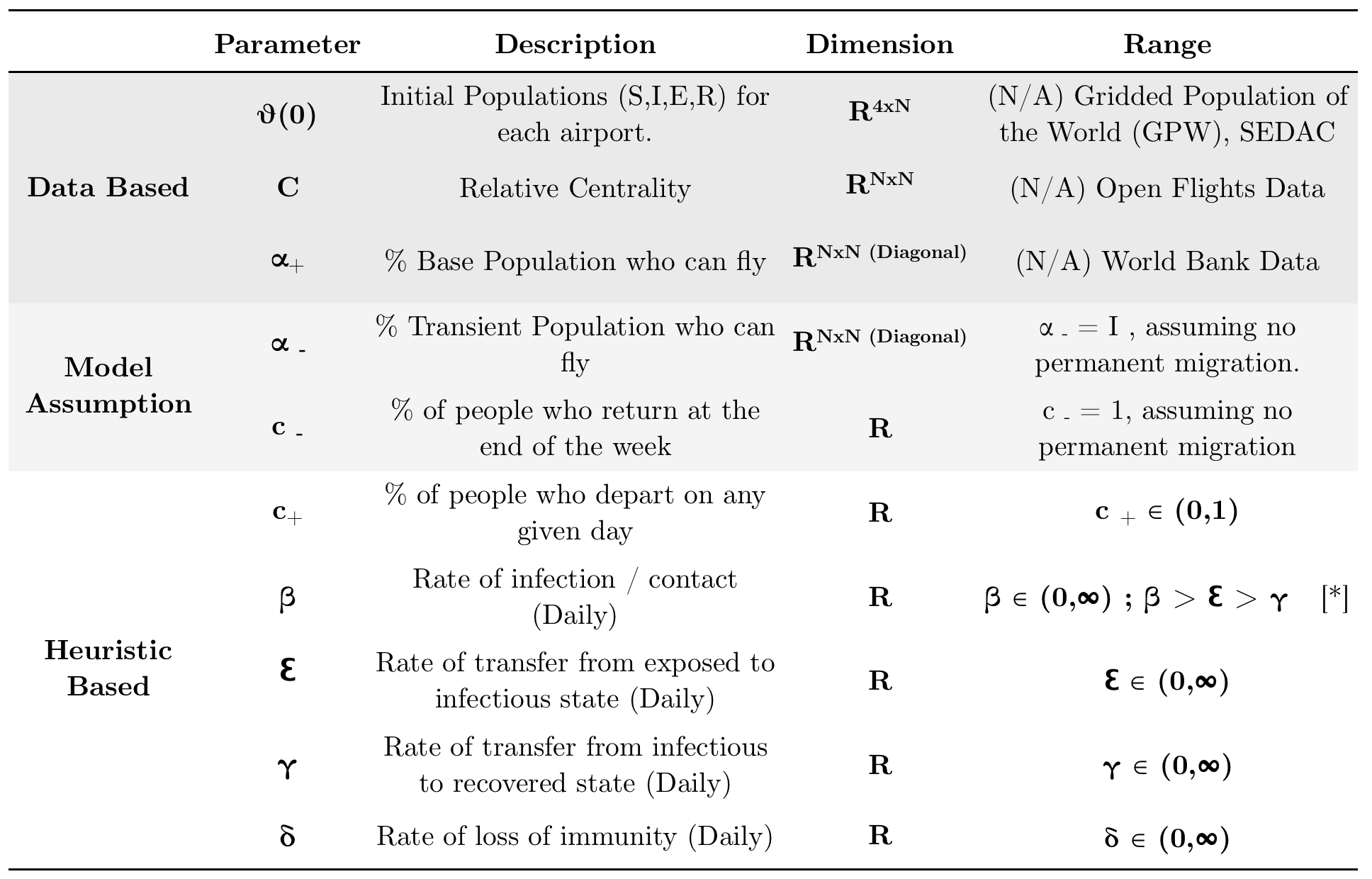}
\end{table}

The international flights network, introduce and analysed in Section \ref{sec:NetworkTopology}, is the foundation for the diffusion part of our model.
From this network of route connections, we aim at constructing a migration matrix connecting more than $3{,}000$ locations.
One fundamental transformation that we apply is the following: let $P$ be the vector of Page-Rank values obtained from earlier analysis and $A$ be the adjacency matrix of flight connections. Then, we define and work with the matrix $C$, which is the relative centrality matrix with elements defined by $C_{ij} = \frac{A_{ij}P_j}{\sum_k A_{ik}P_k}$. The interpretation of this new matrix is that the positions of its non-null entries are identical to those of $A$, except now the edges have been assigned weights based on the relative importance of adjacent nodes. This transformation ensures that the flows of individuals between airports are scaled so that they reflect higher traffic to major airports and less traffic to smaller airports, thus preventing the system from diverging towards unrealistic configurations as the simulation progresses.

With regard to the distribution of the population into a metapopulation structure, we must define the number of individuals that have access to, and are served by, each of the airports. 
This corresponds to estimating the initial state of the system, $\theta$, in that we are estimating the counts $M$ for each airport location.
For this task, we use the gridded population of the world dataset \cite{GPW}, and we assign a value for total population at each airport location. In order to perform this assignment, there is an immediate problem because many airports are often in close proximity of each other (for example some cities are served by multiple airports). We assume that the maximal distance that anyone will travel to reach an airport is $240$km ($60$ km/h * $4$ hours = $240$km). Using this assumption, for each cell of the grid of the population dataset, we search all the airports that are within this radius. Then, we assign a population contribution to each, proportionally to their Page-Rank values. This metric is chosen because it provides a reasonable indication of which airport the travellers will tend to use. Note that this approach will automatically exclude population grid cells which are not within $240$km of any airport. These populations are excluded from the simulation as they are unlikely to be flying regularly, yet we note that this corresponds to less than $3\%$ of the total population.

Finally, we estimate the percentage of each population who can fly $\alpha_{+}$. It is obvious that this percentage will vary across countries, depending on a number of factors, primarily wealth. To find a reasonable value for this model parameter, we acquire passenger estimates by country as supplied by the World Bank and divide these through by the total airport populations for the given country. We then assume the proportion by airport is the same as at country level \footnote{Sometimes this proportion is greater than $1$ particularly for major hubs, where annual traffic through the airport may exceed the size of the local population assigned. In these cases we replace our estimate by the average global proportion, clearly this is imperfect but is likely as accurate as we will be able to obtain given the fact that a more correct information is not readily available.}.

One further fundamental modification that we make regarding $\alpha_{+}$, is that the infectious individuals in the $I$ compartment are not allowed to travel.
This assumption is motivated by the fact that individuals suffering from the disease, especially if symptomatic, may be incapacitated to fly, or would likely be identified as infectious by mandatory testing procedures, hence their contributions to the epidemic would be under some reasonable control.
Still, this assumption does not stop the disease from spreading between locations, since the exposed individuals in each of the $E$ compartments would still be carrying the disease.

The SEIRS model, as well as the graph diffusion model, relies on several strong assumptions that inevitably impact the results.
For clarity, we list and summarise below these assumptions, to highlight the specific features that our approach will exhibit and that should be kept in mind. 
\begin{enumerate}
  \item \textbf{Fully mixed local populations:} within any given node, every member of the population has equal chance of contact, and, thus, equal chance of passing on the disease.
  \item \textbf{Maximal travel distance:} we assume the maximum distance someone will travel is $240$km to get to an airport, and thus anyone who is based outside of all airport radiuses is assumed to be isolated and excluded from the model.
  \item \textbf{Air transit only:} we assume that the only way for the disease to spread between nodes is via air routes and that spread via other means are negligible.
  \item \textbf{Fully mixed wealth:} the proportion of population which may fly is the same for all airports in the same country.
  \item \textbf{Heavy symptoms:} the infectious individuals do not travel.
  \item \textbf{No permanent immigration:} outbound and inbound air traffic volume are balanced for each location. 
  \item \textbf{Universal rates:} we assume that the parameters of the SEIRS model are universal and do not vary between countries.
\end{enumerate}
While it is obvious that each of these assumptions can have a non-negligible effect, we do argue that our simulations can show a remarkable resemblance with a realistic epidemic, such as the recent COVID-19 one.
However, as we will show in the next sections, we consider a variety of different epidemic framework to ensure that our results can be generalised, and to provide a tool that can be used in an arbitrary epidemic setting.

\section{Simulations}\label{sec:Results}
\subsection{Visualisation of Unmitigated Spread}\label{sec:unmitigated}
Now that we have outlined the theory and processes to develop our model, we proceed to simulate and visualise an unmitigated epidemic (i.e. no measures to decrease the infection rate). 
One crucial decision that we need to make regards the calibration of the epidemic parameters.
The previous study by \cite{hou2020effectiveness} used a SEIRS model and estimated $\epsilon = 0.14$ , $\gamma = 0.048$ and $\beta=0.4$, with $\delta = 1/730$ set conservatively as there is uncertainty as to how long recovered individuals will remain immune to COVID-19. 
Since we aim at recreating a realistic epidemic resembling the recent COVID-19 spread, we consider parameter ranges around the above values, and run a sensitivity analysis to explore the various different outcomes.

The SEIRS model parameters that we consider are as follows
\begin{itemize}
 \item[$\cdot$] $\epsilon = 1/7$: individuals remain in $E$ for an average of $7$ days;
 \item[$\cdot$] $\delta = 1/730$: recovered individuals remain immune for an average of two years;
 \item[$\cdot$] $\gamma \in \{1/5,\ 1/25,\ 1/50\}$: three possible scenarios where individuals remain infectious for an average of $5$, $25$ or $50$ days on average;
 \item[$\cdot$] $\beta(t) = 3e^{-\eta t /365}$, for $t=0,1,2,\dots$: each infectious individual has an average of $3$ contacts, however this number is scaled down with an exponential decay as time progresses. This is a reflection of the fact that the response to an epidemic will play a crucial role in slowing down or stopping the disease, and that such response will improve over time, due to a variety of factors including social distancing, testing, better understanding of the disease and treatments.
 \item[$\cdot$] $\eta \in \{0, 2.5, 5, 7.5, 10\}$: the exponential decay  can have different slopes, the value $0$ corresponds to no-slope and thus a constant $\beta$, whereas $10$ gives a dramatic decrease and thus it corresponds to a very effective response.
\end{itemize}
As a \textbf{default configuration}, we set $\eta = 5$, $\epsilon = 1/7$, $\gamma = 1/25$ and $\delta = \frac{1}{730}$.

In terms of initial conditions, we setup the model such that the first cases occur in the Wuhan Airport metapopulation, for similarity with the COVID-19 outbreak \cite{shereen2020covid}. 
In the initial state of the system, all of the metapopulations are in the susceptible compartment, with the exception of Wuhan Airport where we have $0.001\%$ of the local population in the infectious compartment.

As an initial example, we run the model with the default configuration and show a qualitative analysis of the results. 
This simulation corresponds to an unmitigated scenario, because all the airports remain fully operational throughout the simulation, and the only response to the epidemic is the one implied by the parameter $\eta = 5$, i.e. a fairly slow exponential decay of the average number of contacts.

\begin{figure}[htpb]
\centering
\includegraphics[width=0.6\textwidth]{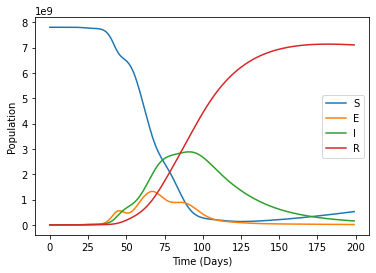}
\caption{Aggregated SEIRS compartments' sizes, for the unmitigated scenario under the default configuration.}
\label{fig:unmitigated_1}
\end{figure}

\begin{figure}[htpb]
\centering
\includegraphics[width=.9\textwidth]{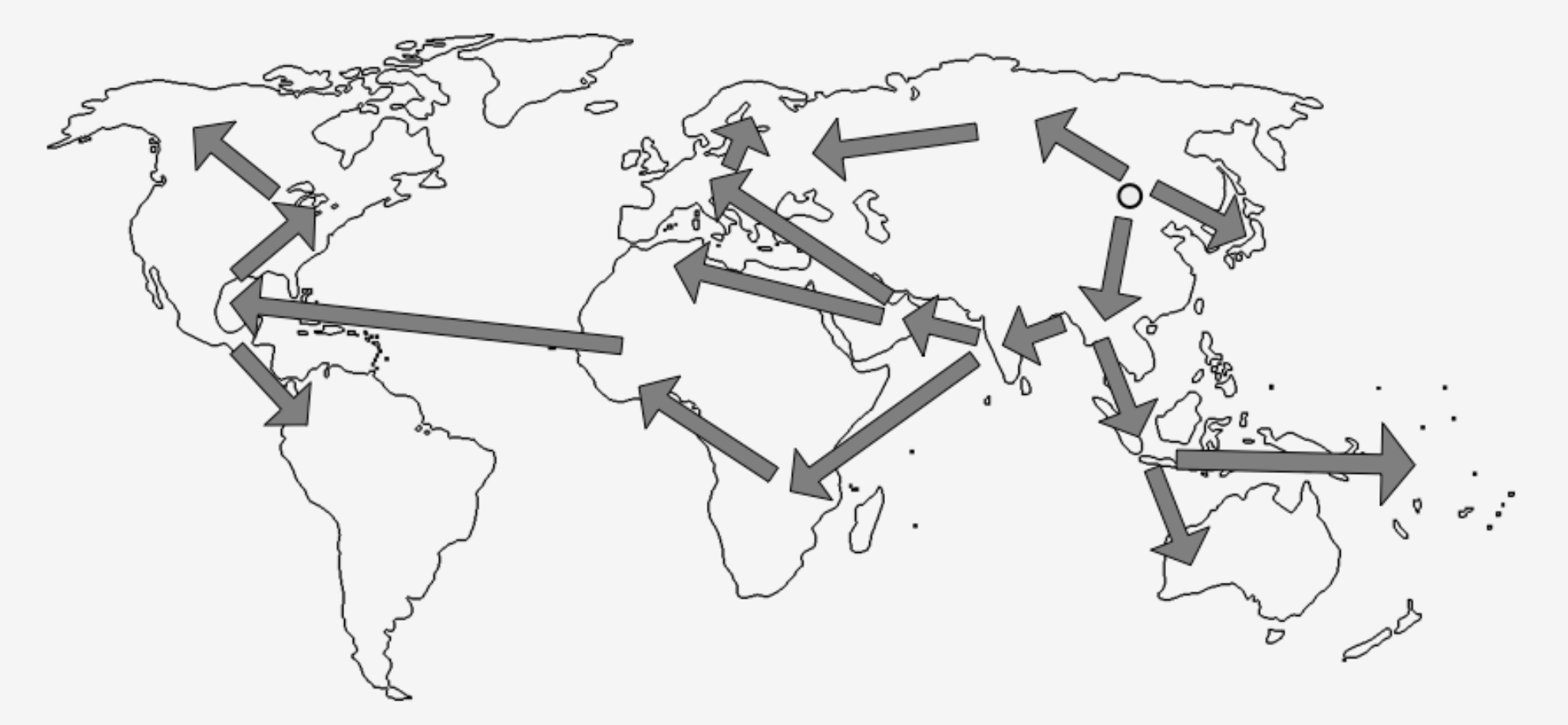}
\caption{Illustrative sketch of the epidemic based on the simulations of Figure \ref{fig:unmitigated_1}.}
\label{fig:Sketch}
\end{figure}

The simulation's results, shown in Figure \ref{fig:unmitigated_1}, are clearly tragic, in that the epidemic easily escalates into a pandemic and affects almost all individuals.
While unrealistic, this worst case scenario example illustrates how airport closure interventions can potentially play a crucial role in slowing the epidemic, but it also allows us to visualise the spatio-temporal spread of the disease.
Figure \ref{fig:Sketch} is an illustrative sketch of the spread through the network which is derived from our simulations. 
There is evidence of a gradual dispersion of the virus across the world starting in China, moving onward into other areas including South East Asia, Japan, Russia, India, South Africa, Middle East. Some of the last places to be infected are the Americas, Nordic Countries, Alaska and Turkey. 

\subsection{Sensitivity Analysis of Key Parameters}
In this section, in order to obtain a better understanding of the impact of these SEIRS parameters on our results, we conduct a sensitivity analysis. 
As per the epidemic parameters considered in Section \ref{sec:unmitigated}, here we run the simulations for the epidemic parameters varying in their respective ranges, for the unmitigated scenario.
This means that the outcome of the epidemic is solely determined by the epidemic parameters, because there are no airport closures.
When extracting the results, we focus on a few key airport locations, and we also consider a benchmark scenario where there is no network structure (i.e. all population is in the same location).

In Table \ref{tab:sim_table_bench} we report the max and the argmax of the $I$ compartment, for each of the locations, and for the various combinations of $\eta$ and $\gamma$.
\begin{table}[!htbp]
\scriptsize
\caption{\textbf{Sensitivity analysis}. The tables on the left column show the time until the peak of $I$ is attained, at that particular location. The tables on the right column show the max value of $I$ (in millions) at that particular location. The simulations were run for $200$ days. Local populations (in millions) served by these airports are $8.34$ for London Heathrow Airport, $3.97$ for Dubai International Airport, $28.39$ for Hong Kong International, $8.08$ for JFK International.}
\begin{minipage}{.5\linewidth}
\centering
\bgroup
\def\arraystretch{1.5}
\begin{tabular}{a*{5}{c}}
\toprule
\rowcolor{white}
\multicolumn{6}{c}{\textbf{London Heathrow Airport:}} \\
\multicolumn{6}{c}{\textbf{Days until peak of infections}} \\
\midrule
\rowcolor{Gray}
\backslashbox{$\gamma$}{$\eta$} & \makebox[1.25em]{0} & \makebox[1.25em]{2.5} & \makebox[1.25em]{5} & \makebox[1.25em]{7.5} & \makebox[1.25em]{10} \\
 0.2 &   57 &   65 &   78 &  105 &  105 \\
0.04 &   60 &   62 &   70 &   80 &  101 \\
0.02 &   65 &   65 &   75 &   85 &  100 \\
\bottomrule
\end{tabular}
\egroup
\vspace{1em}
\end{minipage}
\begin{minipage}{.5\linewidth}
\centering
\bgroup
\def\arraystretch{1.5}
\begin{tabular}{a*{5}{c}}
\toprule
\rowcolor{white}
\multicolumn{6}{c}{\textbf{London Heathrow Airport:}} \\
\multicolumn{6}{c}{\textbf{Peak of infections (millions)}} \\
\midrule
\rowcolor{Gray}
\backslashbox{$\gamma$}{$\eta$} & \makebox[1.25em]{0} & \makebox[1.25em]{2.5} & \makebox[1.25em]{5} & \makebox[1.25em]{7.5} & \makebox[1.25em]{10} \\
 0.2 &  3.1 &  3.0 &  2.6 &  1.2 &  0.0 \\
0.04 &  6.8 &  6.9 &  7.1 &  7.2 &  6.3 \\
0.02 &  8.4 &  8.6 &  8.9 &  9.1 &  9.1 \\
\bottomrule
\end{tabular}
\egroup
\vspace{1em}
\end{minipage}
\begin{minipage}{.5\linewidth}
\centering
\bgroup
\def\arraystretch{1.5}
\begin{tabular}{a*{5}{c}}
\toprule
\rowcolor{white}
\multicolumn{6}{c}{\textbf{Dubai International Airport:}} \\
\multicolumn{6}{c}{\textbf{Days until peak of infections}} \\
\midrule
\rowcolor{Gray}
\backslashbox{$\gamma$}{$\eta$} & \makebox[1.25em]{0} & \makebox[1.25em]{2.5} & \makebox[1.25em]{5} & \makebox[1.25em]{7.5} & \makebox[1.25em]{10} \\
 0.2 &   57 &   65 &   79 &  106 &  105 \\
0.04 &   60 &   65 &   70 &   81 &  105 \\
0.02 &   65 &   70 &   80 &   85 &  105 \\
\bottomrule
\end{tabular}
\egroup
\vspace{1em}
\end{minipage}
\begin{minipage}{.5\linewidth}
\centering
\bgroup
\def\arraystretch{1.5}
\begin{tabular}{a*{5}{c}}
\toprule
\rowcolor{white}
\multicolumn{6}{c}{\textbf{Dubai International Airport:}} \\
\multicolumn{6}{c}{\textbf{Peak of infections (millions)}} \\
\midrule
\rowcolor{Gray}
\backslashbox{$\gamma$}{$\eta$} & \makebox[1.25em]{0} & \makebox[1.25em]{2.5} & \makebox[1.25em]{5} & \makebox[1.25em]{7.5} & \makebox[1.25em]{10} \\
 0.2 &  3.9 &  4.0 &  3.7 &  1.8 &  0.0 \\
0.04 &  8.7 &  9.1 &  9.7 & 10.2 &  9.6 \\
0.02 & 10.9 & 11.6 & 12.3 & 13.3 & 14.0 \\
\bottomrule
\end{tabular}
\egroup
\vspace{1em}
\end{minipage}
\begin{minipage}{.5\linewidth}
\centering
\bgroup
\def\arraystretch{1.5}
\begin{tabular}{a*{5}{c}}
\toprule
\rowcolor{white}
\multicolumn{6}{c}{\textbf{Hong Kong International:}} \\
\multicolumn{6}{c}{\textbf{Days until peak of infections}} \\
\midrule
\rowcolor{Gray}
\backslashbox{$\gamma$}{$\eta$} & \makebox[1.25em]{0} & \makebox[1.25em]{2.5} & \makebox[1.25em]{5} & \makebox[1.25em]{7.5} & \makebox[1.25em]{10} \\
 0.2 &   46 &   51 &   58 &   70 &   87 \\
0.04 &   47 &   50 &   54 &   60 &   68 \\
0.02 &   51 &   55 &   57 &   62 &   70 \\
\bottomrule
\end{tabular}
\egroup
\vspace{1em}
\end{minipage}
\begin{minipage}{.5\linewidth}
\centering
\bgroup
\def\arraystretch{1.5}
\begin{tabular}{a*{5}{c}}
\toprule
\rowcolor{white}
\multicolumn{6}{c}{\textbf{Hong Kong International:}} \\
\multicolumn{6}{c}{\textbf{Peak of infections (millions)}} \\
\midrule
\rowcolor{Gray}
\backslashbox{$\gamma$}{$\eta$} & \makebox[1.25em]{0} & \makebox[1.25em]{2.5} & \makebox[1.25em]{5} & \makebox[1.25em]{7.5} & \makebox[1.25em]{10} \\
 0.2 &  9.4 &  9.0 &  8.2 &  6.3 &  2.1 \\
0.04 & 20.4 & 20.5 & 20.6 & 20.6 & 19.9 \\
0.02 & 24.8 & 25.0 & 25.2 & 25.4 & 25.5 \\
\bottomrule
\end{tabular}
\egroup
\vspace{1em}
\end{minipage}
\begin{minipage}{.5\linewidth}
\centering
\bgroup
\def\arraystretch{1.5}
\begin{tabular}{a*{5}{c}}
\toprule
\rowcolor{white}
\multicolumn{6}{c}{\textbf{JFK International:}} \\
\multicolumn{6}{c}{\textbf{Days until peak of infections}} \\
\midrule
\rowcolor{Gray}
\backslashbox{$\gamma$}{$\eta$} & \makebox[1.25em]{0} & \makebox[1.25em]{2.5} & \makebox[1.25em]{5} & \makebox[1.25em]{7.5} & \makebox[1.25em]{10} \\
 0.2 &   59 &   67 &   81 &  109 &  105 \\
0.04 &   60 &   63 &   70 &   82 &  105 \\
0.02 &   61 &   66 &   75 &   85 &  105 \\
\bottomrule
\end{tabular}
\egroup
\vspace{1em}
\end{minipage}
\begin{minipage}{.5\linewidth}
\centering
\bgroup
\def\arraystretch{1.5}
\begin{tabular}{a*{5}{c}}
\toprule
\rowcolor{white}
\multicolumn{6}{c}{\textbf{JFK International:}} \\
\multicolumn{6}{c}{\textbf{Peak of infections (millions)}} \\
\midrule
\rowcolor{Gray}
\backslashbox{$\gamma$}{$\eta$} & \makebox[1.25em]{0} & \makebox[1.25em]{2.5} & \makebox[1.25em]{5} & \makebox[1.25em]{7.5} & \makebox[1.25em]{10} \\
 0.2 &  2.8 &  2.7 &  2.2 &  0.8 &  0.0 \\
0.04 &  6.1 &  6.2 &  6.2 &  6.1 &  5.0 \\
0.02 &  7.4 &  7.6 &  7.7 &  7.8 &  7.4 \\
\bottomrule
\end{tabular}
\egroup
\vspace{1em}
\end{minipage}
\begin{minipage}{.5\linewidth}
\centering
\bgroup
\def\arraystretch{1.5}
\begin{tabular}{a*{5}{c}}
\toprule
\rowcolor{white}
\multicolumn{6}{c}{\textbf{Benchmark (no network):}} \\
\multicolumn{6}{c}{\textbf{Days until peak of infections}} \\
\midrule
\rowcolor{Gray}
\backslashbox{$\gamma$}{$\eta$} & \makebox[1.25em]{0} & \makebox[1.25em]{2.5} & \makebox[1.25em]{5} & \makebox[1.25em]{7.5} & \makebox[1.25em]{10} \\
 0.2 &   30 &   32 &   34 &   37 &   41 \\
0.04 &   33 &   34 &   35 &   37 &   39 \\
0.02 &   36 &   37 &   38 &   40 &   41 \\
\bottomrule
\end{tabular}
\egroup
\end{minipage}
\begin{minipage}{.5\linewidth}
\centering
\bgroup
\def\arraystretch{1.5}
\begin{tabular}{a*{5}{c}}
\toprule
\rowcolor{white}
\multicolumn{6}{c}{\textbf{Benchmark (no network):}} \\
\multicolumn{6}{c}{\textbf{Peak of infections (millions)}} \\
\midrule
\rowcolor{Gray}
\backslashbox{$\gamma$}{$\eta$} & \makebox[1.25em]{0} & \makebox[1.25em]{2.5} & \makebox[1.25em]{5} & \makebox[1.25em]{7.5} & \makebox[1.25em]{10} \\
 0.2 &  0.7 &  0.6 &  0.6 &  0.6 &  0.5 \\
0.04 &  1.4 &  1.4 &  1.4 &  1.4 &  1.4 \\
0.02 &  1.7 &  1.7 &  1.7 &  1.7 &  1.7 \\
\bottomrule
\end{tabular}
\egroup
\end{minipage}
\label{tab:sim_table_bench}
\end{table}
The max of the $I$ compartment is a crucial measure of the impact of an epidemic, since it is directly related to the stress caused to the local health institutions.
As regards the argmax of the $I$ compartment, this can be used as a measure of the speed of the disease: if the disease progresses slowly, the epidemic curve is flat and it is easier to keep the situation under control until better solutions or treatments are found.

We can see in the results in Table \ref{tab:sim_table_bench} that, as $\eta$ increases, the max of $I$ tends to decrease and the argmax of $I$ tends to increase.
This means that actioning an effective response makes the epidemic curve flatter, in that its peak is reduced and delayed.

The different values of $\gamma$ correspond instead to a different duration of the disease in individuals. 
In most cases, we see that a smaller value of $\gamma$, which signals a longer duration of the disease, causes higher peaks of infections indicating a more difficult situation.
The relation of $\gamma$ with the timing of the peak seems less clear, possibly because if the peak decreases then it might also be reached sooner, but this would not indicate a more difficult situation, in general.

The table also shows that the two parameters jointly affect the results: the impact of $\eta$ on the results is stronger when $\gamma$ is large, and, vice versa, $\eta$ has less of an effect when $\gamma$ is small.
If the duration of the disease is short (large $\gamma$), then we see a relatively low number of cases, which (for most locations) drops down to $0.0$ when $\beta$ decays at a fast rate.
The situation is different when the duration of the disease is long (small $\gamma$): in this case we see more infections overall, and the number of infections does not decrease for faster decays of $\beta$.
In fact, for the major airports that we consider, we observe a counter intuitive scenario where the peak of infections increases when $\beta$ decays at a faster rate.
This is motivated by travelling: the airports that we consider are hubs in the network, meaning that they have a substantial and constant in-flow of exposed individuals, which then switch to infected while staying in the major city.
Clearly this situation does not occur in the no-travelling benchmark framework; however, even in this case, we can notice that the peak of infection is essentially unaffected by the decay rate of $\beta$.

\section{Mitigation Strategies}
In this section we examine potential mitigation strategies in the context of epidemics on international flight networks. 
We introduce several mitigation strategies and test them on our SEIRS model.
The strategies that we consider are the following:
\begin{itemize}
 \item[$\cdot$] \textbf{Nth day rule}: close all airports after a prefixed number of days.
 \item[$\cdot$] \textbf{Threshold rule}: close an airport if the number of infected reaches a prefixed threshold.
 \item[$\cdot$] \textbf{Limited Nth day rule}: sort the airports based on some nodal attribute, and then close a number of them after a prefixed number of days.
 \item[$\cdot$] \textbf{Optimisation approach}: use a genetic algorithm to determine which airport is best to close after a prefixed number of days.
\end{itemize}

In order to evaluate these strategies we must first select some performance metrics. We decide to select metrics which are easy to interpret by policy makers and by the general public, while also being useful in the context of managing hospital intensive care unit capacity and overall impact of the epidemic. Specifically, we will measure the peak number of infections and the total cases (measured as the peak of recovered individuals), as these are transparent and can easily be measured from our simulations.
We will report these values in relation to a baseline, which is provided by the unmitigated scenario of Section \ref{sec:unmitigated}, whereby no airports are closed.

In all simulations we use the default configuration of the SEIRS parameters, but for some of the models we also vary $\eta$ to provide more complete results.
We argue that the results can be generalised and that the same qualitative results are obtained with alternative realistic configurations of the SEIRS parameters.

\subsection{Nth Day Rule}
\begin{figure}[htpb]
	\centering
	\includegraphics[width=0.9\textwidth]{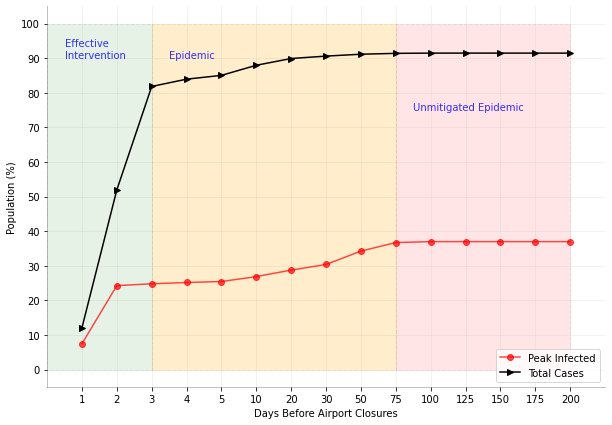}
	\caption{Impact of worldwide permanent airport closures from nth day since first infection.}
	\label{fig:NthDayGraphs}
\end{figure}

The first mitigation strategy that we consider is defined by the permanent closure of air routes from the nth day after the initial outbreak. Our simulations in Figure \ref{fig:NthDayGraphs} demonstrate that closing routes early reduces the peak number of infections, but does not significantly reduce the total number of cases, unless all airports are closed by the end of day $2$. 
While a substantial mitigation can be achieved through airport closures, the time to action these changes is extremely limited, highlighting the importance of timely interventions.
In Table \ref{tab:nth_day_rule}, we further study some of the aspects of the mitigation strategy from a global perspective, in conjunction with different decay values of $\beta$.
\begin{table}[!htbp]
\footnotesize
\centering
\caption{\textbf{Nth day rule}. Mitigation level of the disease measured as a percentage reduction with respect to the baseline which has no airport closures (reported on the last column). The results are reported for various decay parameters (on the rows) and for permanent airport closures happening on different days since the start of the epidemic (over the columns).}
\bgroup
\def\arraystretch{1.5}
\begin{tabular}{a*{5}{c}|c}
\toprule
\rowcolor{white}
\multicolumn{6}{c}{\textbf{\underline{Nth Day Rule:} peak infections}} & \% affected population \\
\multicolumn{6}{c}{\% reduction wrt to worst scenario} & in the worst scenario \\
\rowcolor{Gray}
\backslashbox{$\eta$}{day} & \makebox[1.25em]{\textbf{1}} & \makebox[1.25em]{\textbf{2}} & \makebox[1.25em]{\textbf{3}} & \makebox[1.25em]{\textbf{4}} & \makebox[1.25em]{\textbf{5}} & \makebox[1.25em]{\textbf{200}} \\
\textbf{  0} &    84.1 &    38.3 &    19.1 &    17.5 &    16.5 &    48.3 \\
\textbf{  5} &    79.8 &    34.4 &    33.0 &    32.0 &    31.2 &    37.0 \\
\textbf{ 10} &    60.2 &    58.8 &    57.9 &    57.2 &    56.7 &    15.1 \\
\bottomrule
\end{tabular}\vspace{1em}
\egroup
\bgroup
\def\arraystretch{1.5}
\begin{tabular}{a*{5}{c}|c}
\toprule
\rowcolor{white}
\multicolumn{6}{c}{\textbf{\underline{Nth Day Rule:} total cases}} & \% affected population \\
\multicolumn{6}{c}{\% reduction wrt to worst scenario} & in the worst scenario \\
\rowcolor{Gray}
\backslashbox{$\eta$}{day} & \makebox[1.25em]{\textbf{1}} & \makebox[1.25em]{\textbf{2}} & \makebox[1.25em]{\textbf{3}} & \makebox[1.25em]{\textbf{4}} & \makebox[1.25em]{\textbf{5}} & \makebox[1.25em]{\textbf{200}} \\
\textbf{  0} &    86.9 &    43.3 &     4.8 &     0.9 &     0.4 &    94.5 \\
\textbf{  5} &    86.8 &    43.2 &    10.5 &     8.2 &     7.0 &    91.5 \\
\textbf{ 10} &    77.1 &    76.3 &    75.1 &    73.7 &    72.1 &    51.0 \\
\bottomrule
\end{tabular}
\egroup
\label{tab:nth_day_rule}
\end{table}
We note that closing airports immediately after day $1$ would have a significant impact, reducing the total number of cases by almost $80\%$ for the default configuration. 
This demonstrates the high level of connectivity within the network, with cases of infections proliferating across every continent within just the first $3$ days of the outbreak. 

One fundamental aspect of the epidemic, which is especially apparent in the unmitigated scenario of Table \ref{tab:sim_table_bench}, is that the actual spread of the disease only becomes noticeable after several weeks, when the number of infections starts peaking in some of the locations.
This reflects well how the recent COVID-19 was able to spread across the world unnoticed for months before it was identified and treated as a pandemic. 

Our interpretation of these results is that the graph diffusion model spreads tiny portions of the epidemic in each of the seed's neighbours, thus seeding also these new locations. These ``seeds'' could be interpreted as local patient zeros. 
Then, if the SEIRS parameters guarantee a local escalation of the epidemic, we inevitably observe an increasing number of cases even a long time after the airport closures. 
Once the airports are closed, the escalation of local epidemics become independent of one another and solely dependent on the local epidemic parameters, i.e. on the decay $\eta$.
This is in fact a very positive message, since in these cases the local escalations can be mitigated or even reversed if the number of infectious individuals is small enough, and if $\beta$ decays rapidly.
We can see this clearly in \ref{tab:nth_day_rule}, where faster decay rates lead to much greater mitigation, with smaller numbers of total infections and recoveries.

We also notice that the impact of closures is more relevant when there is no decay, whereas the effect is less noticeable for faster decays. 
This suggests that local interventions aimed at reducing the parameter $\beta$ can play a fundamental role, and potentially replace airport closures.

\subsection{Threshold Rule}
A further modification to the previous results involves dynamically closing airports whenever the the total number of cases exceeds a given percentage of the local population. This differs from the previous method in which we implemented blanket global closures. The results of this new experiment, which are shown in Table \ref{tab:threshold_rule}, indicate that this ``wait and see'' strategy is totally ineffective under our modelling framework. 
\begin{table}[!htbp]
\footnotesize
\centering
\caption{\textbf{Threshold rule}. Mitigation level of the disease measured as a percentage reduction with respect to the baseline which has no airport closures (reported on the last column). The results are reported for various decay parameters (on the rows) and for airport closures happening at different threshold levels (over the columns).}
\bgroup
\def\arraystretch{1.5}
\begin{tabular}{a*{5}{c}|c}
\toprule
\rowcolor{white}
\multicolumn{6}{c}{\textbf{\underline{Threshold Rule:} peak infections}} & \% affected population \\
\multicolumn{6}{c}{\% reduction wrt to worst scenario} & in the worst scenario \\
\rowcolor{Gray}
\backslashbox{$\eta$}{thresh} & \makebox[1.25em]{\textbf{1e-7}} & \makebox[1.25em]{\textbf{1e-5}} & \makebox[1.25em]{\textbf{1e-3}} & \makebox[1.25em]{\textbf{1e-2}} & \makebox[1.25em]{\textbf{1e-1}} & \makebox[1.25em]{\textbf{no threshold}} \\
\textbf{  0} &     2.2 &     1.5 &     1.0 &     0.6 &     0.3 &    48.3 \\
\textbf{  5} &    15.5 &    10.3 &     5.5 &     3.6 &     1.9 &    37.0 \\
\textbf{ 10} &    56.8 &    52.2 &    32.8 &    19.8 &    10.1 &    15.1 \\
\bottomrule
\end{tabular}\vspace{1em}
\egroup
\bgroup
\def\arraystretch{1.5}
\begin{tabular}{a*{5}{c}|c}
\toprule
\rowcolor{white}
\multicolumn{6}{c}{\textbf{\underline{Threshold Rule:} total cases}} & \% affected population \\
\multicolumn{6}{c}{\% reduction wrt to worst scenario} & in the worst scenario \\
\rowcolor{Gray}
\backslashbox{$\eta$}{thresh} & \makebox[1.25em]{\textbf{1e-7}} & \makebox[1.25em]{\textbf{1e-5}} & \makebox[1.25em]{\textbf{1e-3}} & \makebox[1.25em]{\textbf{1e-2}} & \makebox[1.25em]{\textbf{1e-1}} & \makebox[1.25em]{\textbf{no threshold}} \\
\textbf{  0} &     0.0 &     0.0 &     0.0 &     0.0 &     0.0 &    94.5 \\
\textbf{  5} &     0.9 &     0.5 &     0.2 &     0.1 &     0.0 &    91.5 \\
\textbf{ 10} &    65.8 &    48.4 &    26.8 &    16.3 &     8.5 &    51.0 \\
\bottomrule
\end{tabular}
\egroup
\label{tab:threshold_rule}
\end{table}
Even considering a highly unrealistic version of this strategy in where we suppose it is possible to detect cases up to a fineness of $1$ in every $10$ million people, it is already too late to close airports, whereby mitigation is mostly irrelevant unless the decay is really fast.

\subsection{Limited Nth Day Rule}
In the previous results we have shown that it is far more effective to close airports preemptively than it is to wait on some threshold level on infections to be attained within the local population. However, one could argue that it is impractical to close all airports globally, both from a economic and political point of view. This motivates the two following strategies that we propose, where we proceed to examine what performance we can achieve by only closing a subset of key airports, where we can carefully choose the particular subset using different strategies.
Here, we rank the airports using several metrics: population, Page-Rank and betweenness; then, we propose to close a number of airports following the order given by one of these ranking. 
The resulting strategy is similar to the nth day rule, however in this case we selectively close fewer airports based on their metric score.
The main argument and motivation behind this limited nth day rule is that closing all airports will inevitably carry a substantial damage, so we are interested in checking whether good results can also be achieved by closing fewer airports.

Crucially, we observe in Tables \ref{tab:limited_population_rule}, \ref{tab:limited_pagerank_rule} and \ref{tab:limited_betweenness_rule} that even closing just the top $1\%$ of airports we still obtain significantly greater reductions than the threshold rule. 
\begin{table}[!htbp]
\footnotesize
\centering
\caption{\textbf{Limited nth day rule by local population}. Mitigation level of the disease measured as a percentage reduction with respect to the baseline which has no airport closures (reported on the last column). The results are reported for various levels of airport closures happening on different days since the start of the epidemic (over the columns).}
\bgroup
\def\arraystretch{1.5}
\begin{tabular}{a*{5}{c}|c}
\toprule
\rowcolor{white}
\multicolumn{6}{c}{\textbf{\underline{Nth Day Rule by population:} peak infections}} & \% affected population \\
\multicolumn{6}{c}{\% reduction wrt to worst scenario} & in the worst scenario \\
\rowcolor{Gray}
\backslashbox{closed airports}{day} & \makebox[1.25em]{\textbf{1}} & \makebox[1.25em]{\textbf{2}} & \makebox[1.25em]{\textbf{3}} & \makebox[1.25em]{\textbf{4}} & \makebox[1.25em]{\textbf{5}} & \makebox[1.25em]{\textbf{200}} \\
\textbf{   1\%} &    30.1 &     9.4 &     8.4 &     7.8 &     7.5 &    37.0 \\
\textbf{   5\%} &    64.0 &    21.1 &    20.1 &    19.4 &    19.0 &    37.0 \\
\textbf{  10\%} &    79.7 &    29.6 &    28.2 &    27.2 &    26.5 &    37.0 \\
\textbf{  25\%} &    79.8 &    33.7 &    32.3 &    31.3 &    30.5 &    37.0 \\
\textbf{ 100\%} &    79.8 &    34.4 &    33.0 &    32.0 &    31.2 &    37.0 \\
\bottomrule
\end{tabular}\vspace{1em}
\egroup
\bgroup
\def\arraystretch{1.5}
\begin{tabular}{a*{5}{c}|c}
\toprule
\rowcolor{white}
\multicolumn{6}{c}{\textbf{\underline{Nth Day Rule by population:} total cases}} & \% affected population \\
\multicolumn{6}{c}{\% reduction wrt to worst scenario} & in the worst scenario \\
\rowcolor{Gray}
\backslashbox{closed airports}{day} & \makebox[1.25em]{\textbf{1}} & \makebox[1.25em]{\textbf{2}} & \makebox[1.25em]{\textbf{3}} & \makebox[1.25em]{\textbf{4}} & \makebox[1.25em]{\textbf{5}} & \makebox[1.25em]{\textbf{200}} \\
\textbf{   1\%} &    24.2 &     9.5 &     2.9 &     2.1 &     1.7 &    91.5 \\
\textbf{   5\%} &    51.7 &    21.1 &     6.1 &     4.6 &     3.8 &    91.5 \\
\textbf{  10\%} &    72.8 &    28.1 &     8.2 &     6.3 &     5.4 &    91.5 \\
\textbf{  25\%} &    86.8 &    40.5 &     9.9 &     7.7 &     6.6 &    91.5 \\
\textbf{ 100\%} &    86.8 &    43.2 &    10.5 &     8.2 &     7.0 &    91.5 \\
\bottomrule
\end{tabular}
\egroup
\label{tab:limited_population_rule}
\end{table}
\begin{table}[!htbp]
\footnotesize
\centering
\caption{\textbf{Limited nth day rule by Page-Rank centrality}. Mitigation level of the disease measured as a percentage reduction with respect to the baseline which has no airport closures (reported on the last column). The results are reported for various levels of airport closures happening on different days since the start of the epidemic (over the columns).}
\bgroup
\def\arraystretch{1.5}
\begin{tabular}{a*{5}{c}|c}
\toprule
\rowcolor{white}
\multicolumn{6}{c}{\textbf{\underline{Nth Day Rule by Page-Rank:} peak infections}} & \% affected population \\
\multicolumn{6}{c}{\% reduction wrt to worst scenario} & in the worst scenario \\
\rowcolor{Gray}
\backslashbox{closed airports}{day} & \makebox[1.25em]{\textbf{1}} & \makebox[1.25em]{\textbf{2}} & \makebox[1.25em]{\textbf{3}} & \makebox[1.25em]{\textbf{4}} & \makebox[1.25em]{\textbf{5}} & \makebox[1.25em]{\textbf{200}} \\
\textbf{   1\%} &    19.3 &    11.0 &    10.0 &     9.4 &     8.9 &    37.0 \\
\textbf{   5\%} &    74.0 &    31.7 &    29.8 &    28.6 &    27.8 &    37.0 \\
\textbf{  10\%} &    79.5 &    34.1 &    32.5 &    31.4 &    30.5 &    37.0 \\
\textbf{  25\%} &    79.8 &    34.3 &    32.9 &    31.9 &    31.1 &    37.0 \\
\textbf{ 100\%} &    79.8 &    34.4 &    33.0 &    32.0 &    31.2 &    37.0 \\
\bottomrule
\end{tabular}\vspace{1em}
\egroup
\bgroup
\def\arraystretch{1.5}
\begin{tabular}{a*{5}{c}|c}
\toprule
\rowcolor{white}
\multicolumn{6}{c}{\textbf{\underline{Nth Day Rule by Page-Rank:} total cases}} & \% affected population \\
\multicolumn{6}{c}{\% reduction wrt to worst scenario} & in the worst scenario \\
\rowcolor{Gray}
\backslashbox{closed airports}{day} & \makebox[1.25em]{\textbf{1}} & \makebox[1.25em]{\textbf{2}} & \makebox[1.25em]{\textbf{3}} & \makebox[1.25em]{\textbf{4}} & \makebox[1.25em]{\textbf{5}} & \makebox[1.25em]{\textbf{200}} \\
\textbf{   1\%} &     5.1 &     1.5 &     0.8 &     0.7 &     0.6 &    91.5 \\
\textbf{   5\%} &    52.2 &    13.2 &     4.0 &     3.0 &     2.5 &    91.5 \\
\textbf{  10\%} &    81.5 &    31.4 &     8.3 &     6.2 &     5.1 &    91.5 \\
\textbf{  25\%} &    86.7 &    42.1 &     9.9 &     7.5 &     6.3 &    91.5 \\
\textbf{ 100\%} &    86.8 &    43.2 &    10.5 &     8.2 &     7.0 &    91.5 \\
\bottomrule
\end{tabular}
\egroup
\label{tab:limited_pagerank_rule}
\end{table}
\begin{table}[!htbp]
\footnotesize
\centering
\caption{\textbf{Limited nth day rule by betweenness centrality}. Mitigation level of the disease measured as a percentage reduction with respect to the baseline which has no airport closures (reported on the last column). The results are reported for various levels of airport closures happening on different days since the start of the epidemic (over the columns).}
\bgroup
\def\arraystretch{1.5}
\begin{tabular}{a*{5}{c}|c}
\toprule
\rowcolor{white}
\multicolumn{6}{c}{\textbf{\underline{Nth Day Rule by betweenness:} peak infections}} & \% affected population \\
\multicolumn{6}{c}{\% reduction wrt to worst scenario} & in the worst scenario \\
\rowcolor{Gray}
\backslashbox{closed airports}{day} & \makebox[1.25em]{\textbf{1}} & \makebox[1.25em]{\textbf{2}} & \makebox[1.25em]{\textbf{3}} & \makebox[1.25em]{\textbf{4}} & \makebox[1.25em]{\textbf{5}} & \makebox[1.25em]{\textbf{200}} \\
\textbf{   1\%} &    17.8 &     9.4 &     8.7 &     8.2 &     7.8 &    37.0 \\
\textbf{   5\%} &    70.9 &    36.0 &    33.8 &    32.3 &    31.2 &    37.0 \\
\textbf{  10\%} &    74.6 &    36.2 &    34.3 &    33.0 &    32.0 &    37.0 \\
\textbf{  25\%} &    79.7 &    34.9 &    33.3 &    32.2 &    31.4 &    37.0 \\
\textbf{ 100\%} &    79.8 &    34.4 &    33.0 &    32.0 &    31.2 &    37.0 \\
\bottomrule
\end{tabular}\vspace{1em}
\egroup
\bgroup
\def\arraystretch{1.5}
\begin{tabular}{a*{5}{c}|c}
\toprule
\rowcolor{white}
\multicolumn{6}{c}{\textbf{\underline{Nth Day Rule by betweenness:} total cases}} & \% affected population \\
\multicolumn{6}{c}{\% reduction wrt to worst scenario} & in the worst scenario \\
\rowcolor{Gray}
\backslashbox{closed airports}{day} & \makebox[1.25em]{\textbf{1}} & \makebox[1.25em]{\textbf{2}} & \makebox[1.25em]{\textbf{3}} & \makebox[1.25em]{\textbf{4}} & \makebox[1.25em]{\textbf{5}} & \makebox[1.25em]{\textbf{200}} \\
\textbf{   1\%} &     5.6 &     1.5 &     0.7 &     0.6 &     0.5 &    91.5 \\
\textbf{   5\%} &    55.9 &    19.2 &     6.5 &     4.9 &     4.0 &    91.5 \\
\textbf{  10\%} &    78.2 &    29.3 &     8.7 &     6.7 &     5.6 &    91.5 \\
\textbf{  25\%} &    84.8 &    39.5 &    10.3 &     8.1 &     6.9 &    91.5 \\
\textbf{ 100\%} &    86.8 &    43.2 &    10.5 &     8.2 &     7.0 &    91.5 \\
\bottomrule
\end{tabular}
\egroup
\label{tab:limited_betweenness_rule}
\end{table}
This further emphasises the point that early intervention is far more important in the network than attempting to detect when certain infection thresholds have been breached. We also note that the fall-off in the strategy's performance is quite non linear both across the rows or columns. 

With regard to comparisons between the different metrics, it appears that ranking the airports by local population is most effective for first day closures, however the two other centrality based procedures tend to perform better when airports are closed after day $1$.

\subsection{Optimisation Approach}
In section \ref{sec:Results} we explored several mitigation strategies, ultimately finding greatest flexibility and mitigation in the limited nth day rule. 
This strongly suggests that we could find a very effective airport closure strategy that will achieve excellent mitigation results, without totally disrupting the network connectivity. 
In this section, we employ a genetic algorithm (GA) to search for the optimal combination of airport closures that maximises the utility of the nth Day Rule. GAs, in their simplest form, operate on binary strings called chromosomes which are an encoding of the parameters of interest, commonly referred to as genes. Any particular instance of a chromosome has a genotype which refers to a specific string of bits each with values $1$ or $0$ representing a particular gene’s allele. Once the problem is formulated within this framework, the GA procedure is as follows:

\begin{enumerate}
	\item Define a fitness function $F(X)$ which evaluates the optimality of a given genotype.
	
	\item Initialise a population of chromosomes with randomly assigned genotypes.
	
	\item Evaluate the fitness of all members of the population. Individuals will then be selected for breeding at a frequency proportional to their fitness (this emulates the ``survival of the fittest'' mechanism).
	
	\item In a process known as \textit{crossover}, pairs of \textit{selected} genotypes are split uniformly at some locus along the chromosome and then \textit{recombined}, to form new chromosomes. 
	
	\item \textit{Mutations} are then applied at random to alleles of the recombined chromosomes (simply by bit flipping) in order to prevent irrevocable loss of any characteristic.
	
	\item The process (3 - 5) is then repeated so that the fittest members of the population are selected.
\end{enumerate}

For a more detailed explanation of the entire process we refer to \cite{spall2005introduction}. 
We opt to use a genetic algorithm for this problem since the solutions of our combinatorial problem can be easily represented as binary strings. By contrast, problems of this nature are ill suited to classical optimisation methods such as Gradient Descent, as it is not possible to analytically compute gradients and our search space is too large for approximation methods. Additionally a fitness function can be easily defined from the metrics that we have previously described. In the following paragraph, we provide more details regarding  the setup of the GA into our framework.

\paragraph{GA details.}
We define the key information required to formulate our problem within the GA framework as follows:

\begin{itemize}
 \item \textbf{Chromosome}: instead of optimising \textit{which airports to close on each day}, we choose to optimise \textit{which countries should close their airports on each day}. This reduces the GA search space by $15$ times which is highly desirable for convergence and global optimum. This re-framing is also reasonable because it reflects the likely course of action that we would observe in epidemics, where decisions are often taken at country level.
 
 We define the chromosome as a $195$-bit string where $1$ in the i-th entry indicates that the i-th country's airports are closed, and vice versa $0$s indicate closed airports.
 
 \item \textbf{Fitness function}: in designing the fitness function we seek to balance the economic impacts of airport closures along with the reduction in peak infections and peak recoveries, relative to an unmitigated epidemic. Since we have no prior biases between the following $3$ criteria:
 \begin{equation*}
  \begin{split}
   T &= \mbox{$\%$ in reduction of peak infections} \\
   P &= \mbox{$\%$ in reduction of peak recoveries} \\
   A &= \mbox{$\%$ airports remaining open}
  \end{split}
 \end{equation*}
we aggregate them into the objective functions as follows:
\begin{equation*}
 F(X) = T(X) * P(X) * \sin\left( \frac{\pi}{2} * A(X)\right)
\end{equation*}
where $X$ represents a simulated SEIRS environment that is obtained for a given chromosome.
\end{itemize}

In order to evaluate the fitness function, we must compute the values of $T$ and $P$ (computing $A$ is trivial). This clearly involves inputting the parameters encoded in the chromosomes genotype into our simulation developed in previous sections. We perform this by first using a lookup table to convert between the $195$-bit country closures string specified in the genotype to a $3{,}425$ bit string airport closures vector required by our model. Next we set to zero the rows and columns of the closed airports within the adjacency matrix at the appropriate time steps (to disconnect an airport from the network). Finally we proceed to run the algorithm for $200$ days, which is sufficient to capture the first wave of the epidemic under all conditions.

\subsection{Optimisation Results and Comparisons}
The results of the algorithm are shown in Table \ref{tab:ga_results} in the usual format for consistency.
\begin{table}[!htbp]
\footnotesize
\centering
\caption{\textbf{Genetic algorithm strategy}. Mitigation level of the disease measured as a percentage reduction with respect to the baseline which has no airport closures (reported on the last column). The results are reported for permanent airport closures happening on different days since the start of the epidemic (over the columns). Different airports are closed on different days.}
\bgroup
\def\arraystretch{1.5}
\begin{tabular}{a*{5}{c}|c}
\toprule
\rowcolor{white}
\multicolumn{6}{c}{\textbf{\underline{GA Rule:}}} & \% affected population \\
\multicolumn{6}{c}{\% reduction wrt to worst scenario} & in the worst scenario \\
\rowcolor{Gray}
\backslashbox{measure}{day} & \makebox[1.25em]{\textbf{1}} & \makebox[1.25em]{\textbf{2}} & \makebox[1.25em]{\textbf{3}} & \makebox[1.25em]{\textbf{4}} & \makebox[1.25em]{\textbf{5}} & \makebox[1.25em]{\textbf{200}} \\
max(I) &    79.8 &    36.9 &    29.0 &    29.8 &    31.1 &    37.0 \\
max(R) &    86.8 &    34.1 &     8.9 &     7.2 &     5.6 &    91.5 \\
\bottomrule
\end{tabular}\vspace{1em}
\egroup
\label{tab:ga_results}
\end{table}
\begin{table}[!htbp]
\footnotesize
\centering
\caption{\textbf{Genetic algorithm strategy}. Number of countries that close they airports on each day, according to the GA's solutions. The total number of countries is $220$.}
\bgroup
\def\arraystretch{1.5}
\begin{tabular}{cc}
\toprule
\multicolumn{2}{c}{\textbf{GA strategy}} \\
\midrule
Day & Number of \\
 & closed countries\\
\midrule
1 & 83 \\
2 & 113 \\
3 & 88 \\
4 & 109 \\
5 & 103 \\
\bottomrule
\end{tabular}\vspace{1em}
\egroup
\label{tab:ga_results_closed}
\end{table}
Similarly to the other methods, we see a substantial difference between a day $1$ closure and a day $5$ closure, both with respect to the peak of infections and peak of recoveries.
The mitigation effect, measured as a percentage reduction of the number of cases with respect to the unmitigated scenario, is approximately at the same level as the one obtained with the limited nth day rule.
Differently from the limited nth day rule, the GA approach does not consider airport closures but country closures, which is clearly more realistic but also less efficient.
However, we note that the objective function can be calculated in all of the solutions, and it is maximised by the GA solution, for each day.

\begin{figure}[htpb]
	\begin{subfigure}{0.5\textwidth}
		\centering
		\includegraphics[width=\textwidth]{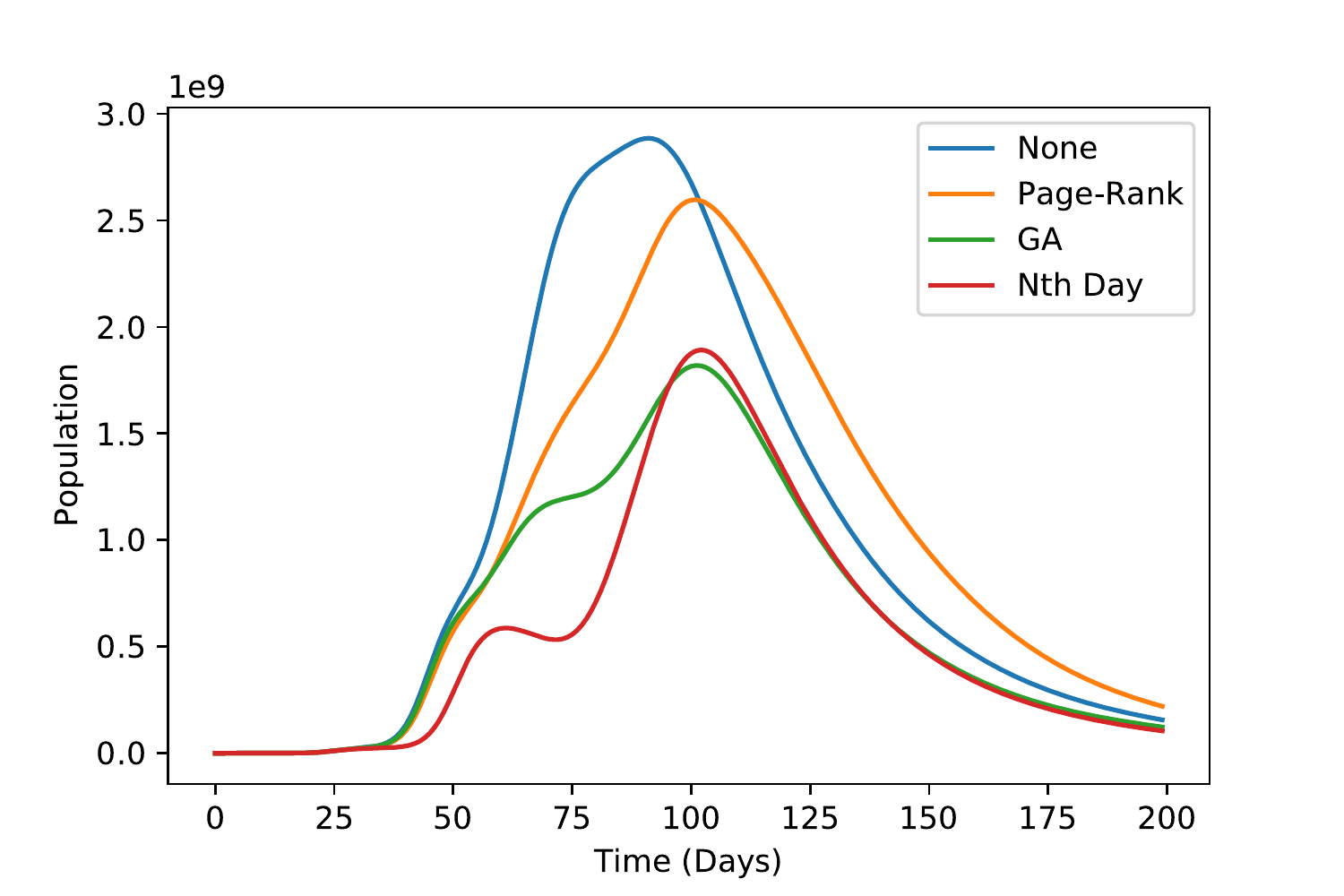}
		\label{fig:ga_compare_infected_2}
		\caption{$I$ compartment for day $2$ closures.}
	\end{subfigure}%
	\begin{subfigure}{0.5\textwidth}
		\centering
		\includegraphics[width=\textwidth]{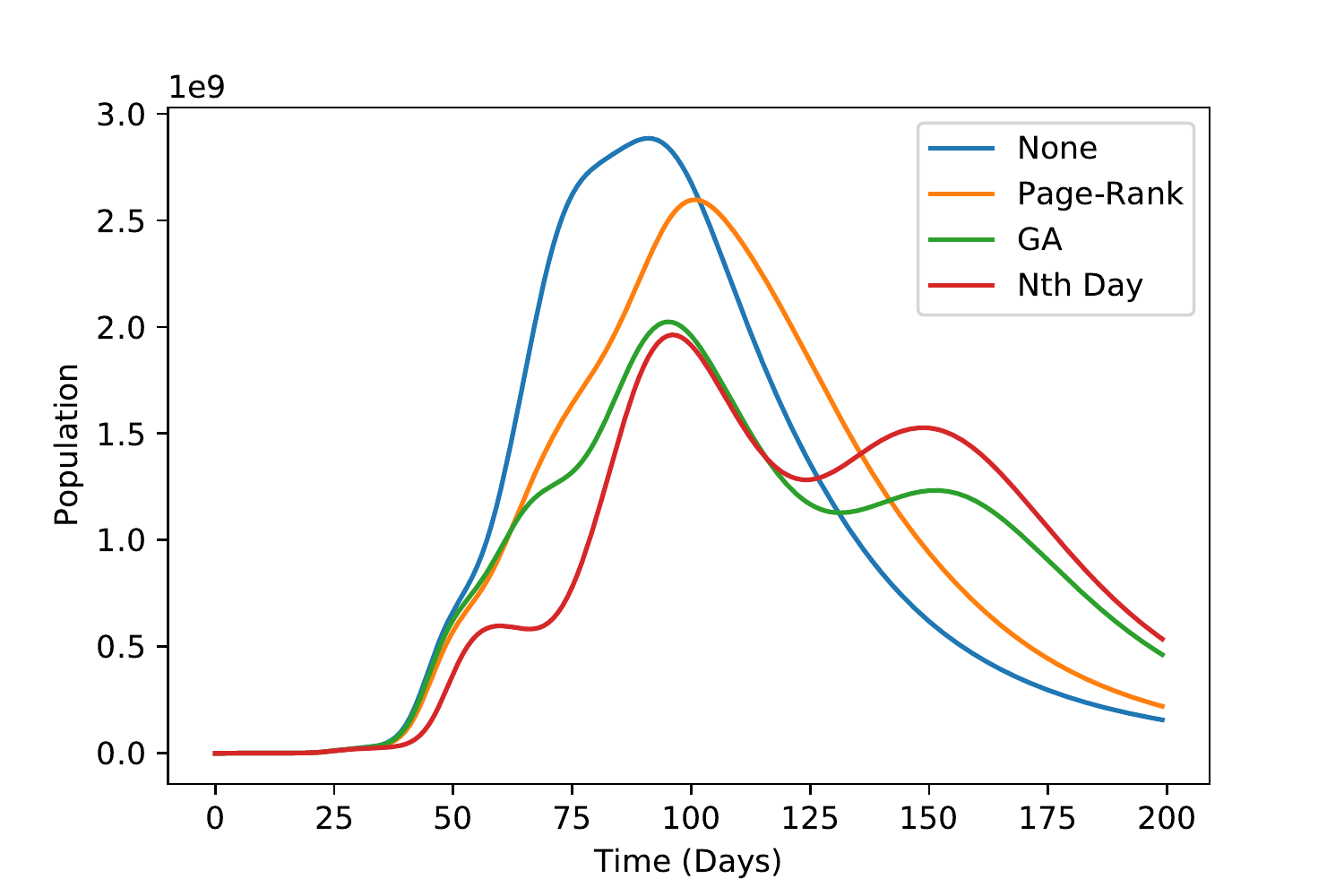}
		\label{fig:ga_compare_infected_4}
		\caption{$I$ compartment for day $4$ closures.}
	\end{subfigure}
	\caption{A comparison of GA strategy versus naive strategies and total inaction.}
	\label{fig:ga_compare_2}
\end{figure}

\begin{figure}[htpb]
	\begin{subfigure}{0.5\textwidth}
		\centering
		\includegraphics[width=\textwidth]{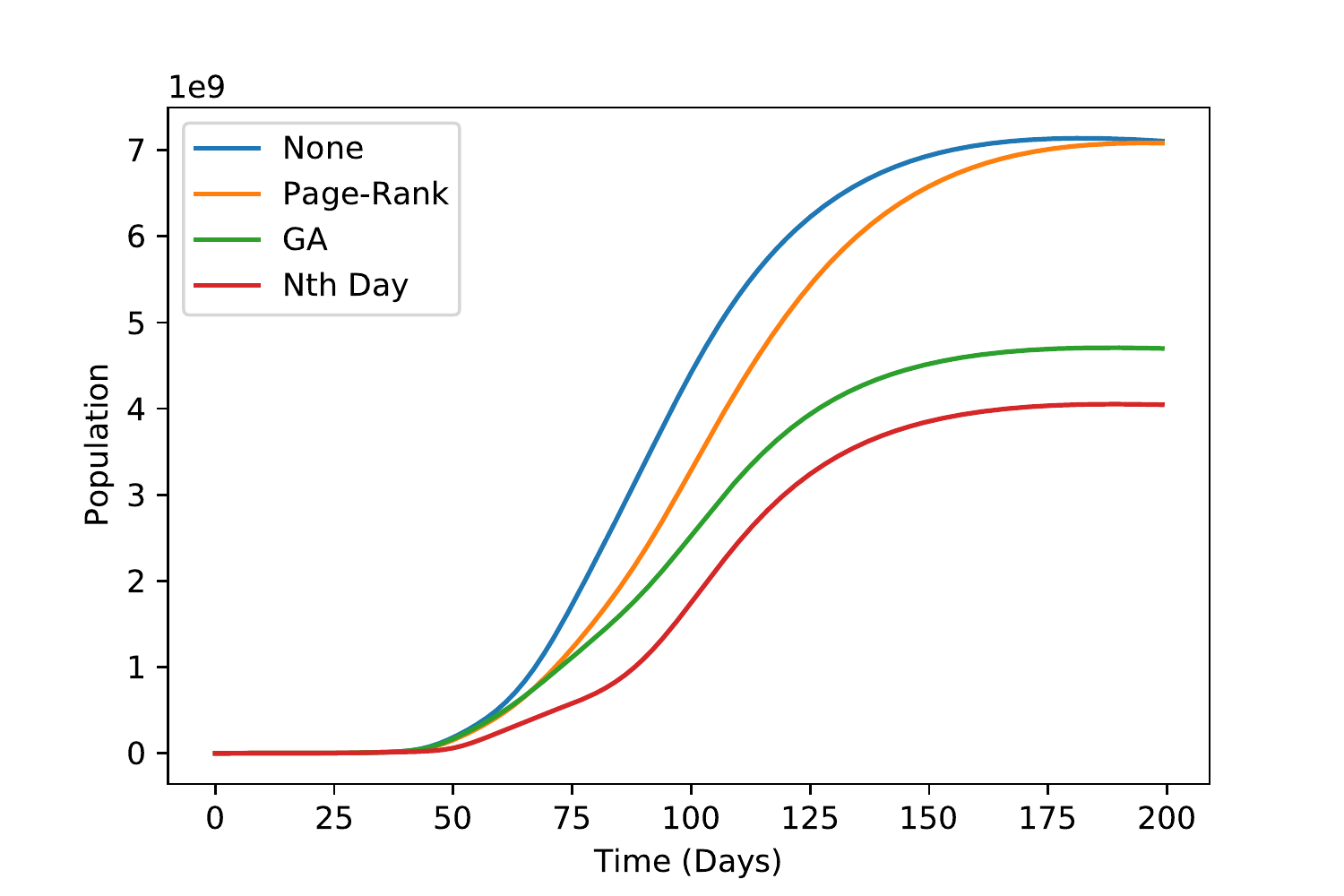}
		\label{fig:ga_compare_recovered_2}
		\caption{$R$ compartment for day $2$ closures.}
	\end{subfigure}%
	\begin{subfigure}{0.5\textwidth}
		\centering
		\includegraphics[width=\textwidth]{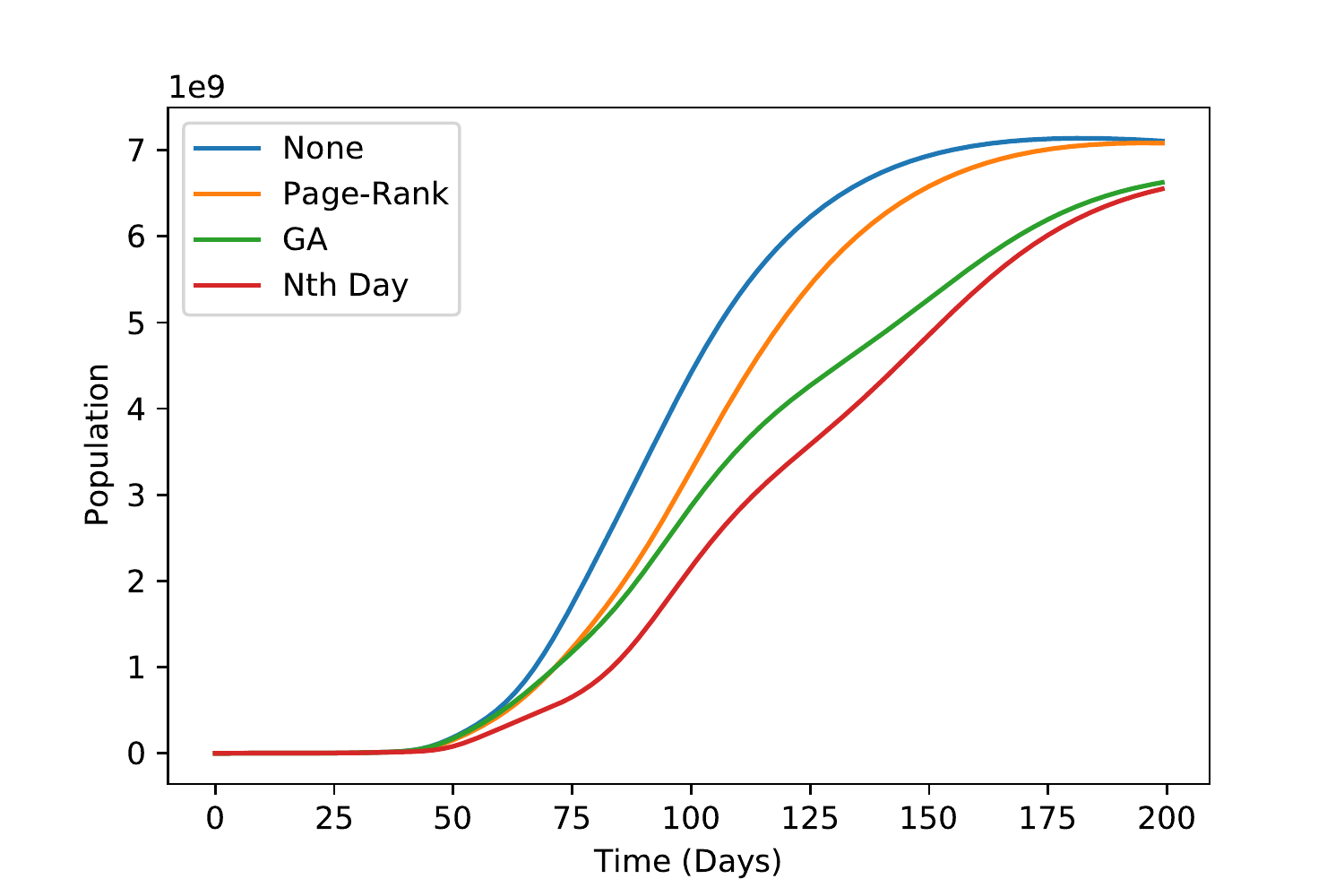}
		\label{fig:ga_compare_recovered_4}
		\caption{$R$ compartment for day $4$ closures.}
	\end{subfigure}
	\caption{\textbf{GA results}. A comparison of SEIRS compartments for the GA strategy versus the other strategies considered. The limited nth day rule uses Page-Rank centrality and closes $1\%$ of the airports.}
	\label{fig:ga_compare_4}
\end{figure}

We offer a more detailed overview of the comparisons between the different mitigation strategies in Figures \ref{fig:ga_compare_2} and \ref{fig:ga_compare_4}.
In these figures we represent the evolution of the two main compartments of interest, $I$ and $R$.
We notice that the closure strategies affect these compartments in a rather non-trivial way, in that, at the global level, they can delay the epidemic and fragment it into various waves (i.e. peaks of infections).

In Figure \ref{fig:ga_compare_2}, the line labeled as \texttt{None} refers to the unmitigated scenario: we see that in this case the epidemic peaks very quickly and that the $I$ compartment reaches the highest value across all methods.
This is the worst scenario that we observe, when compared to the other strategies.

For day $2$ closures, we see that the GA strategy strictly dominates the centrality-based strategy because the infectious cases are fewer at all times.
The nth day rule, which refers to the closure of all airports from the nth day, naturally gives the best results because this also corresponds to the strictest of measures.

The peak of infections is in fact higher than that of the GA strategy: this is perfectly reasonable because the different strategies not only affect the peaks but also the distribution of infections over time.
For day $4$ closures, the results are not as clear, however we argue that also in this case the GA algorithm attains better results than the Page-Rank strategy, getting close to the nth day rule but without as many closures.

Figure \ref{fig:ga_compare_4} highlights in a more clear way the same message, providing a neat ranking of the strategies.
For both day $2$ and day $4$ closures, we can see that the GA achieves excellent results without having to resort to full closures; on the other hand, the centrality based strategy provides little improvement over the unmitigated scenario.

\begin{figure}[htpb]
	\centering
	\includegraphics[width=0.495\textwidth]{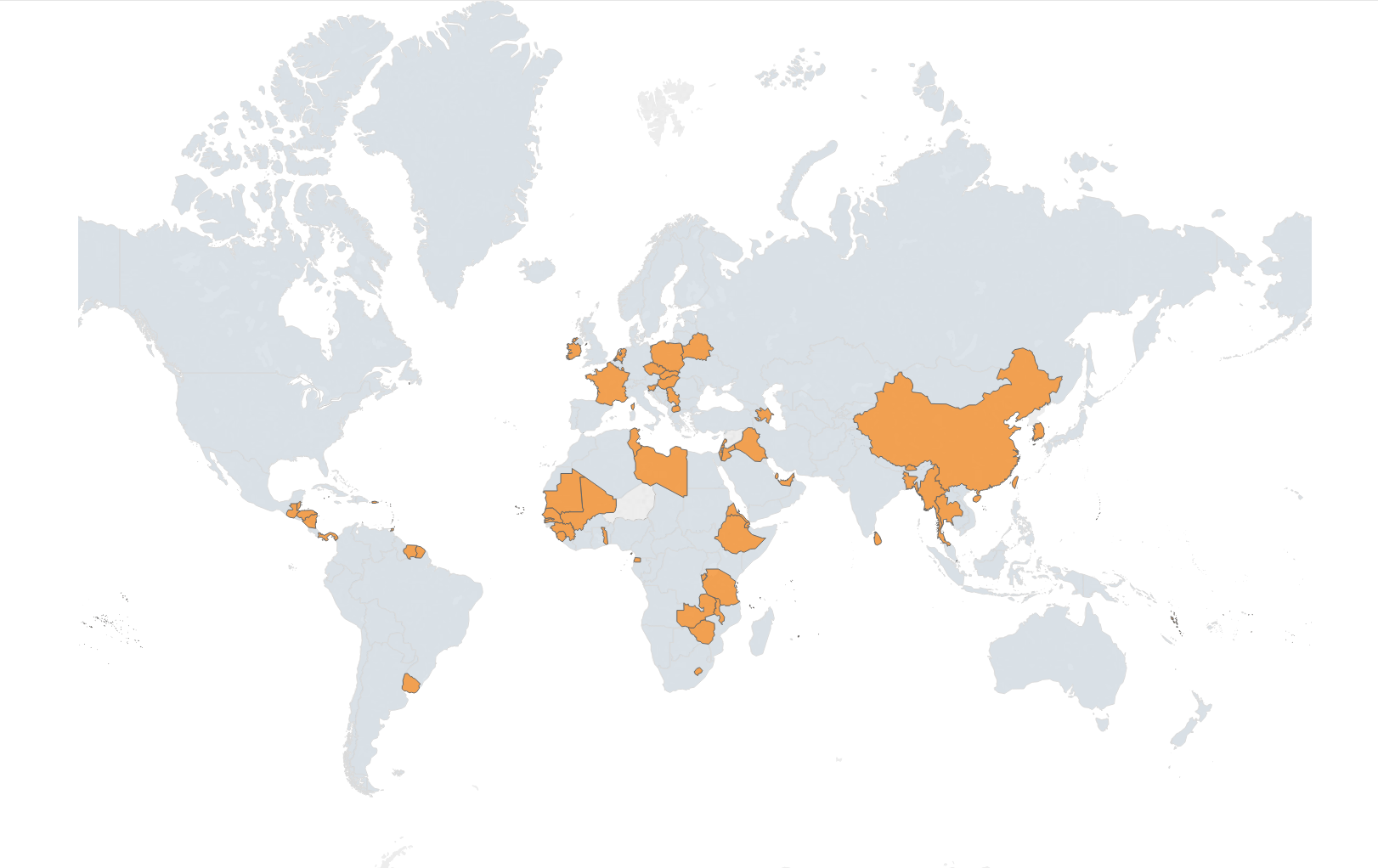}
	\includegraphics[width=0.495\textwidth]{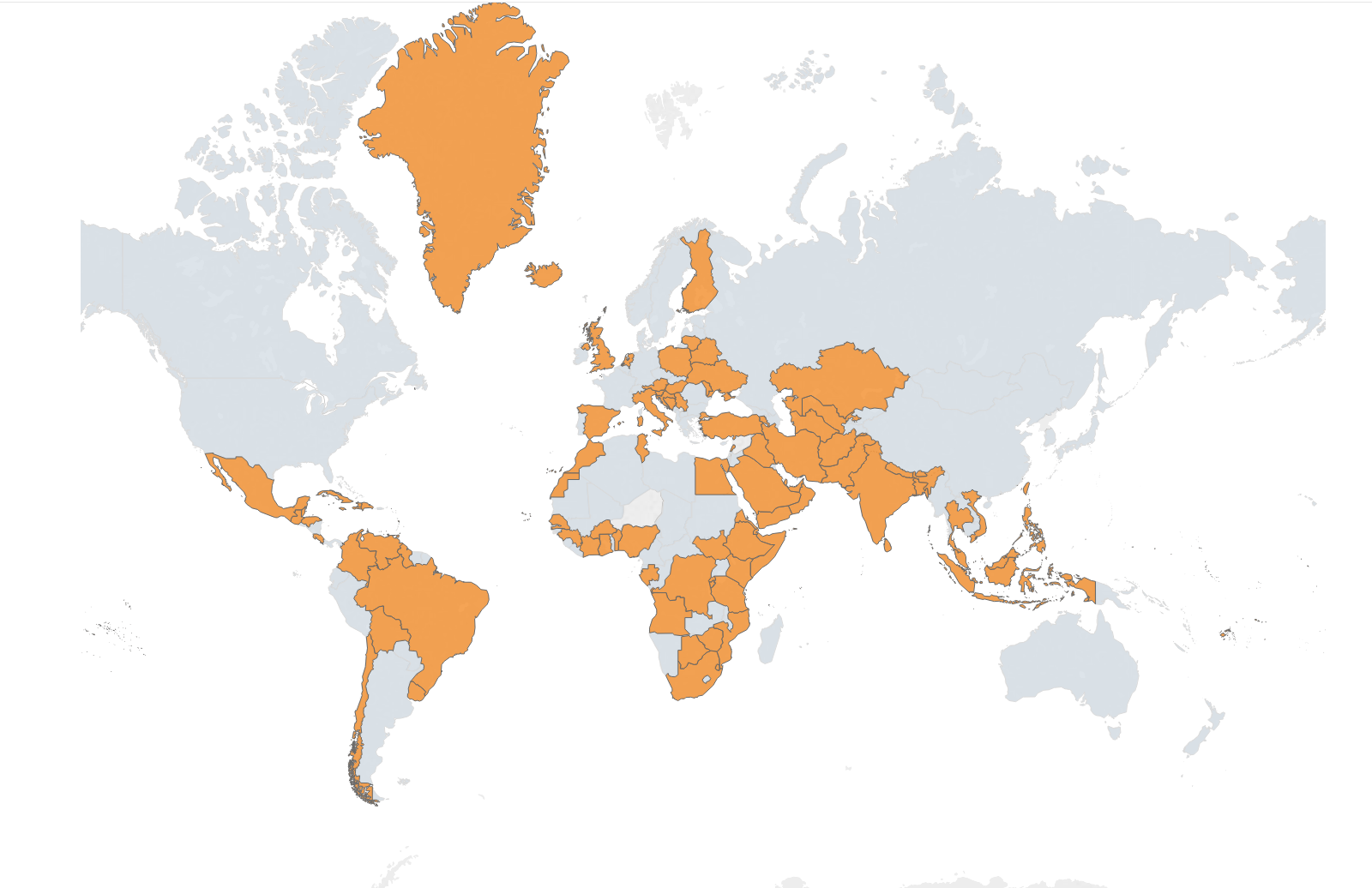}
	\includegraphics[width=0.495\textwidth]{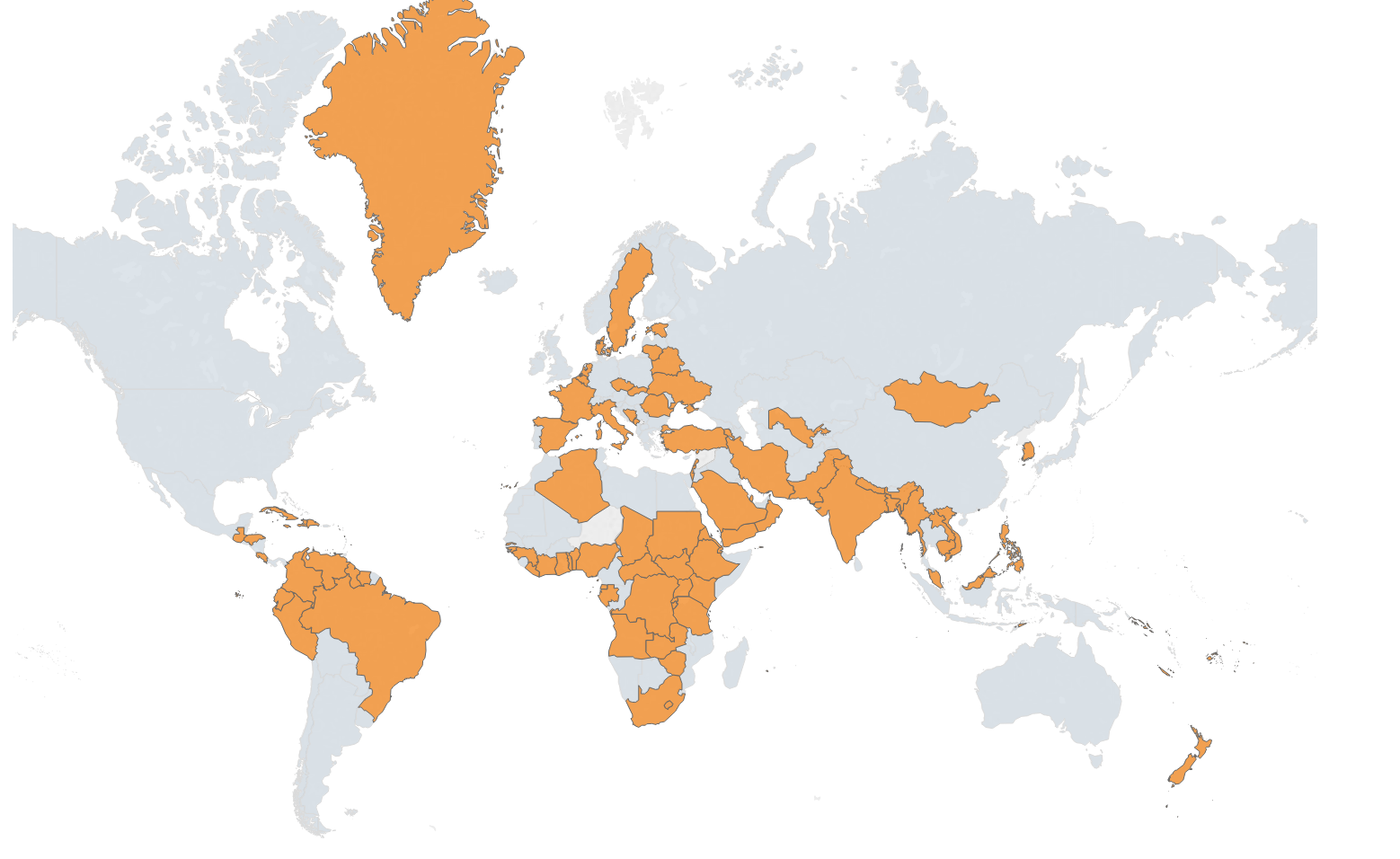}
	\includegraphics[width=0.495\textwidth]{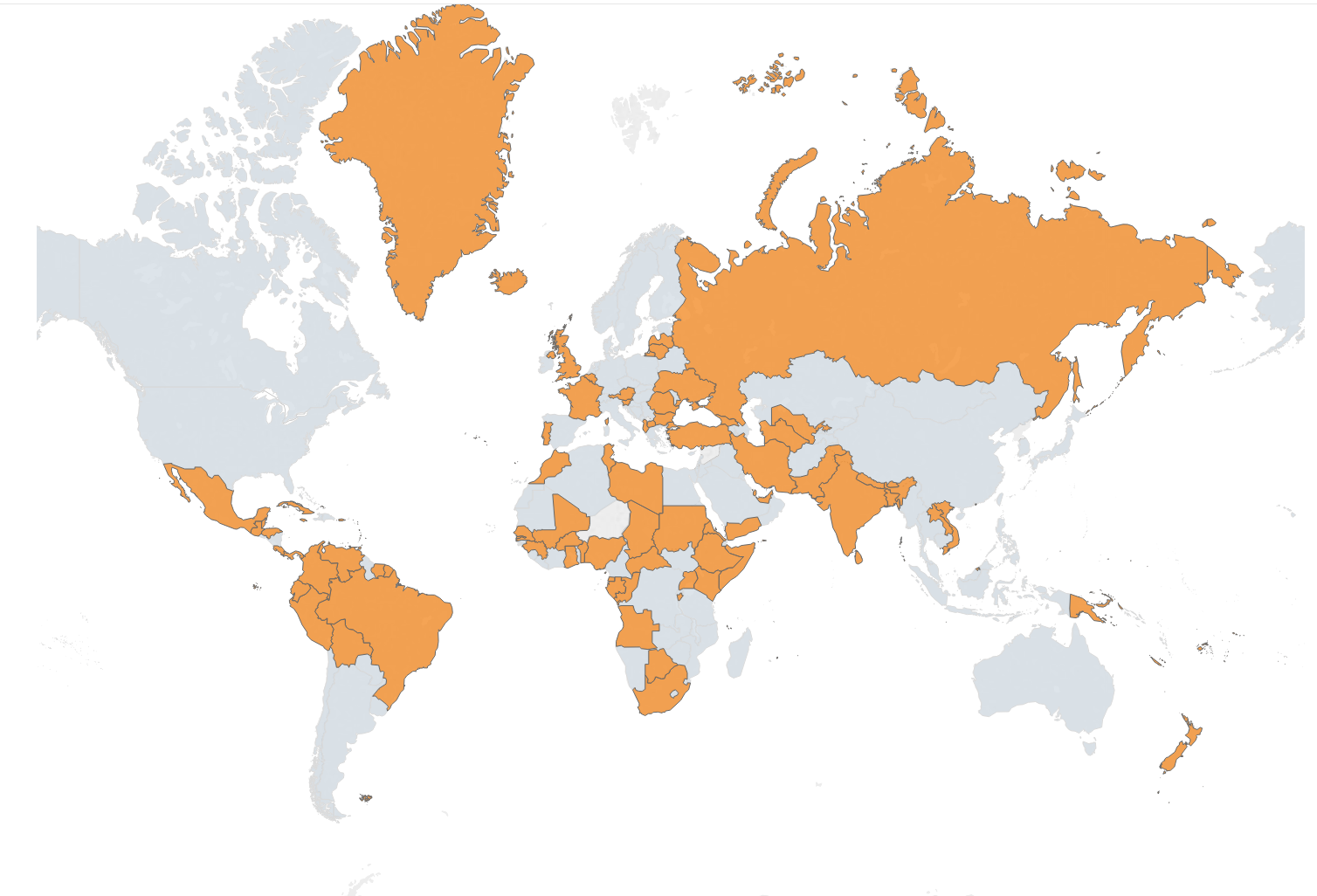}
	\caption{\textbf{GA solutions} for days $1$ to $4$. The countries in dark blue are the ones that should be closed on day $1$ (top-left), day $2$ (top-right), day $3$ (bottom-left) and day $4$ (bottom-right).}
	\label{fig:GAByDay}
\end{figure}

Our results suggests that the GA learns to leverage the network structure via closures in such a way as to accelerate the initial infection rate, but achieving a lower point of equilibrium. To gain further intuition into the GA behaviour, it is best to look at Figure \ref{fig:GAByDay} which exhibits the evolution of the GA strategy as we alter the day at which closures occur. 
What we observe is that it is initially optimal to close China and certain other countries such as France which are very well connected to China via air routes. However as governments delay airport closures, we find that the GA shifts focus away from China and starts to establish ``fire-breaks'' in other regions such as the Middle East, South America, Europe and Africa.

\begin{figure}[htpb]
    \centering
		\includegraphics[width=0.495\textwidth]{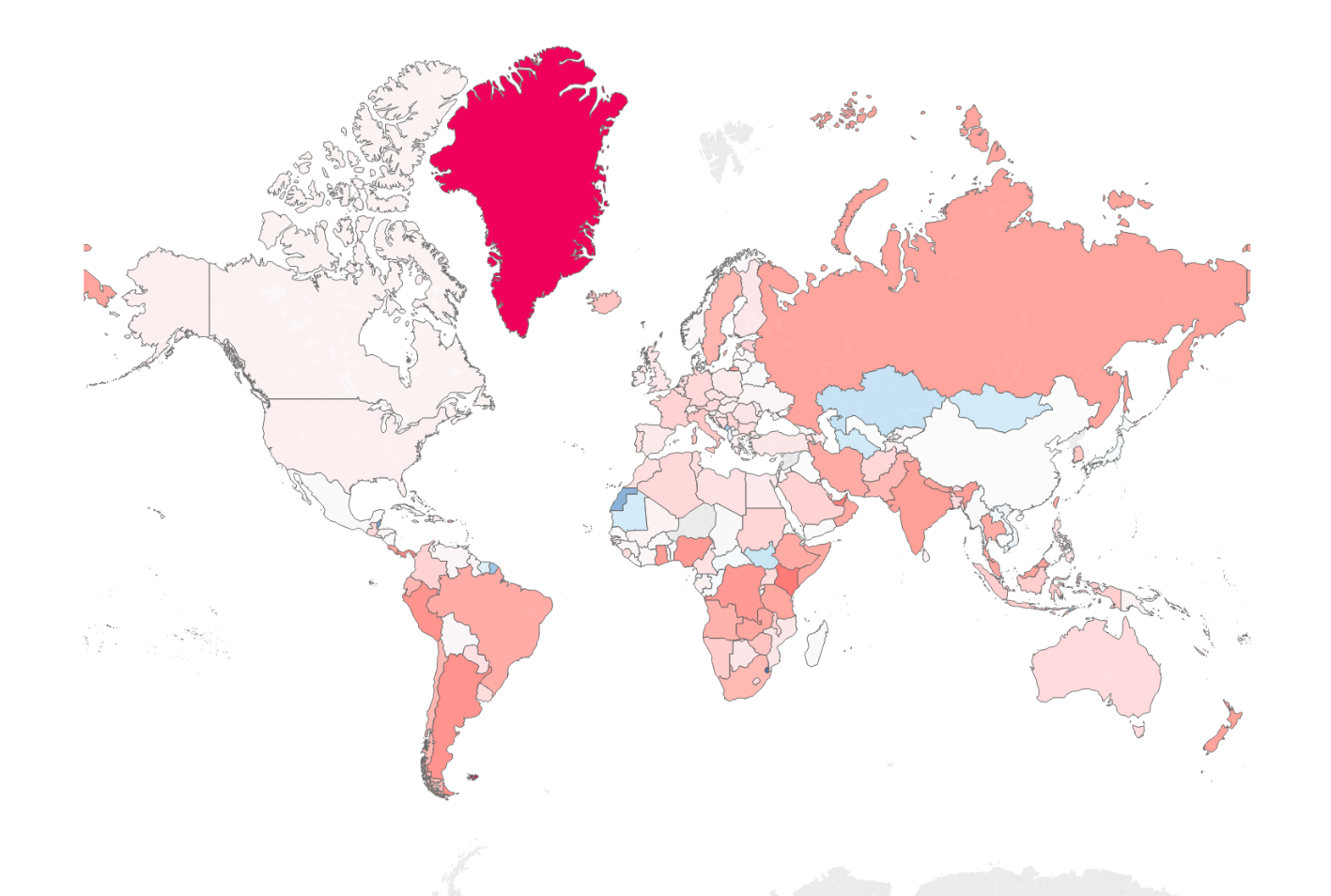}
		\includegraphics[width=0.495\textwidth]{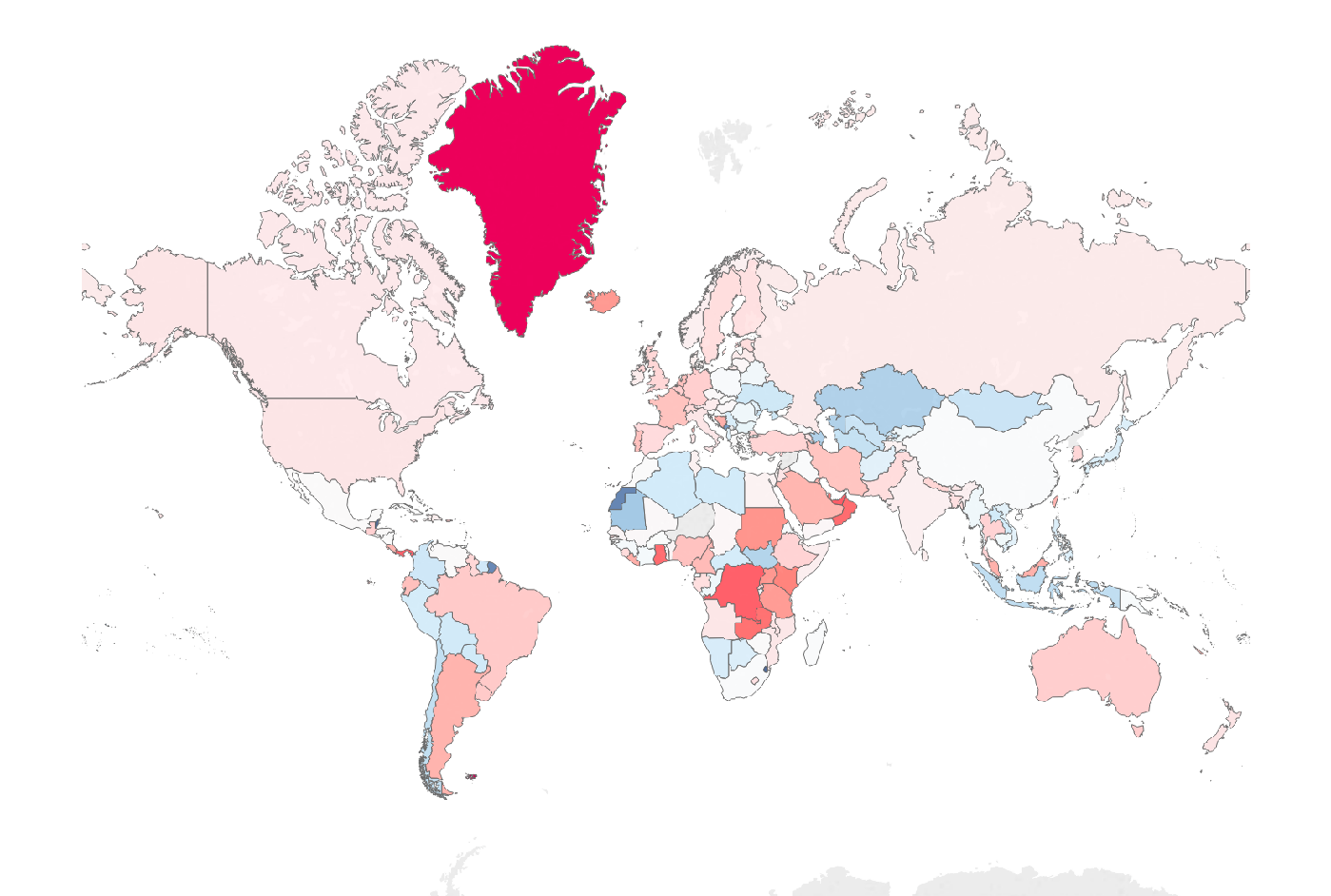}
	\caption{\textbf{GA method}. Cases reductions achieved by the GA solution for day $3$ closures, with respect to the unmitigated scenario. The peak of infections is shown on the left panel, whereas the right panel shows the peak of recoveries. The colours of countries fade from red (GA strategy is favourable) to blue (unmitigated strategy is favourable).}
	\label{fig:GAversus_1}
\end{figure}

\begin{figure}[htpb]
    \centering
    \includegraphics[width=0.495\textwidth]{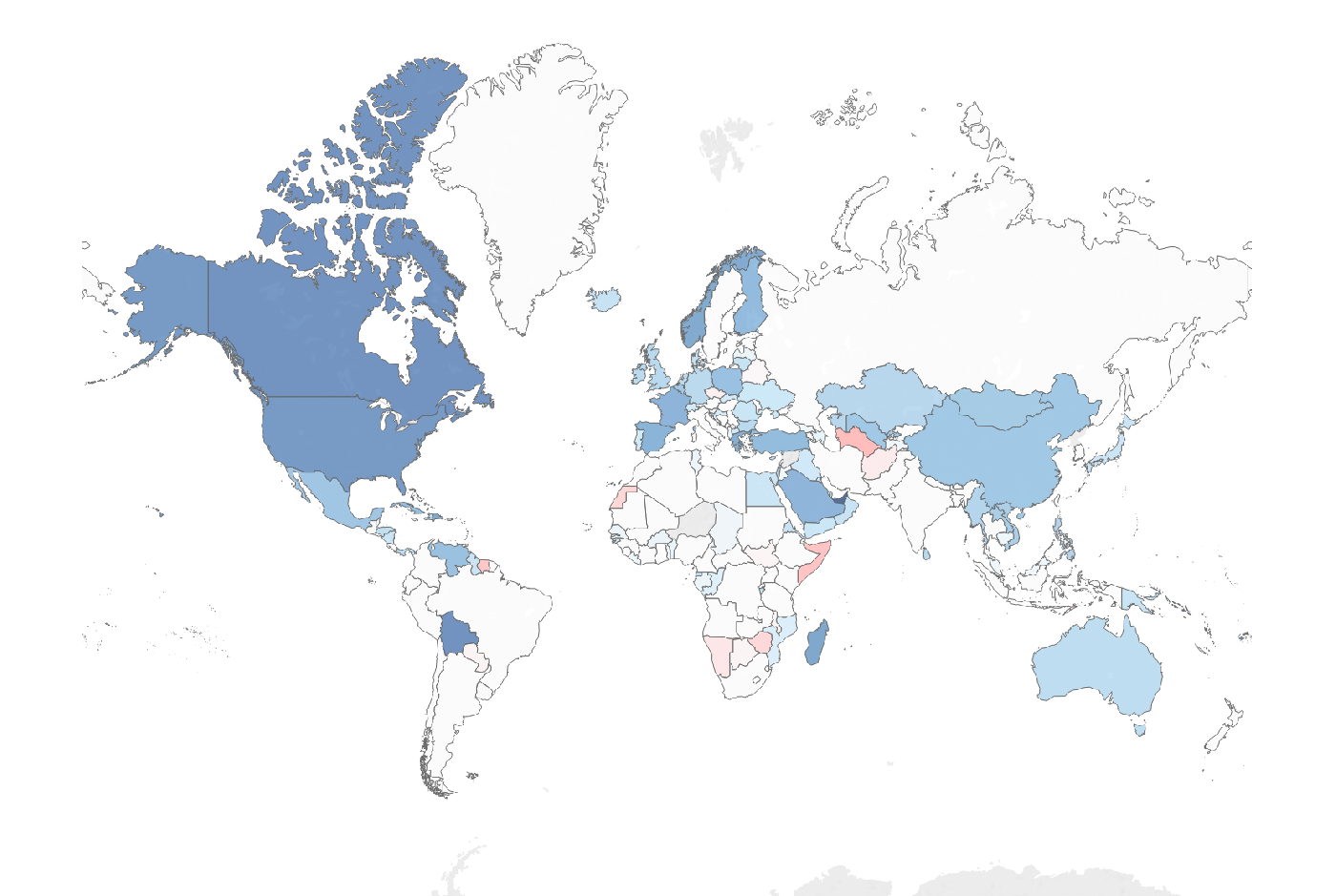}
    \includegraphics[width=0.495\textwidth]{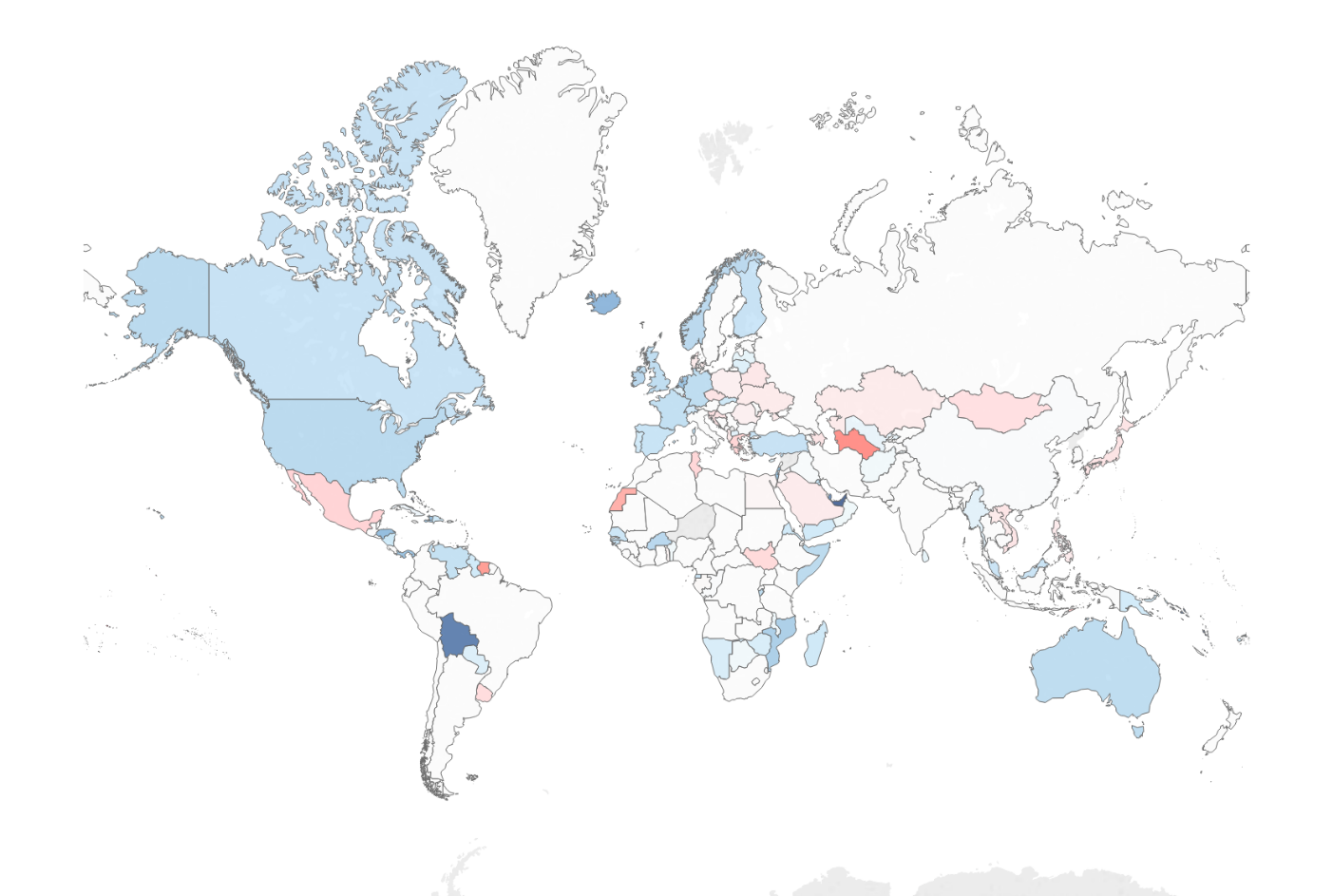}
	\caption{\textbf{GA method}. Cases reductions achieved by the GA solution for day $3$ closures, with respect to the nth day rule. The peak of infections is shown on the left panel, whereas the right panel shows the peak of recoveries. The colours of countries fade from red (GA strategy is favourable) to blue (nth day rule is favourable).}
	\label{fig:GAversus_2}
\end{figure}

In Figures \ref{fig:GAversus_1} and \ref{fig:GAversus_2}, we focus on day $3$ closures and we examine the percentage change in peak infections and peak recoveries under the GA strategy, compared with the unmitigated strategy (i.e. the worst strategy) and nth day rule (i.e. the best strategy), respectively.  
Despite less than half of the airports being closed under the GA strategy, the vast majority of countries see a reduction in peak infections and peak recoveries relative to the unmitigated case.
By contrast, fewer countries see a reduction in peak infections and peak recoveries relative to the nth day rule. 
We note that the countries that are closed in day $3$ roughly correspond to the countries that will see the best reductions for the number of cases.
In particular, we highlight that some highly connected countries such as China and the United States, do not benefit much from the day $3$ GA closure strategy and see little reductions in the number of infections and recoveries (see Figure \ref{fig:GAversus_1}).
However, these countries would benefit more from a close-all strategy, as per the nth day rule (see Figure \ref{fig:GAversus_2}).
Our results provide evidence of the effectiveness of the GA approach over other naive strategies, offering a useful tool can be used to obtain a model-based perspective in this decision problem.

\section{Conclusion}
In this paper, we have shown that human mobility infrastructure networks are highly complex and highly resilient to node removals. This has important consequences in epidemiology, since the high connectivity facilitates the spread of pandemic diseases to a worldwide scale. These observations motivated us to seek and test non-trivial airport closure strategies that could maintain a good network connectivity, while slowing down the spread of a potential epidemic. 

The main contribution of our work is that we illustrate a useful methodology which can be used to study epidemics, predict and analyse possible realistic scenarios, and, thus, support decision making with regard to crucial interventions that could help in confining the epidemic.
We provide an application of our method to a variety of scenarios which, to some extent, also resemble recent COVID-19 epidemic. Our results do not directly relate to COVID-19, rather we take a more general approach and explore a number of realistic circumstances. Our tool is designed in a flexible way and can accommodate a variety of epidemic settings.

The framework that we propose is based on a metapopulation SEIRS model, where the subpopulations are located in the airports' locations, and they are composed of the individuals living in those nearby areas. Our simulations allowed us to explore and study a variety of realistic scenarios to understand the dynamics of the spread of the disease. We considered several different airport closure strategies, and we compared them using measures extracted from our simulations.

One main message that arises from our analysis is that, due to travelling, the disease is seeded in many locations worldwide with impressive speed. If conditions are met for isolated local escalations of the epidemic, then most countries will be hit by the seeded disease after a variable delay, regardless of any late interventions on travelling restrictions.

Our findings suggest that the first week of dispersion of the disease through the network is a critical time period for effective intervention, however interventions in the network, such as airport closures, still provide some reductions to peak infections and total cases if implemented after the initial week. Furthermore, we show that policies which reduce community spread can be combined with our proposed airport closure strategies to provide greater benefits than if either policy had been used separately. 

Finally, we explored the application of an optimisation approach to identify optimal airport closures within the critical first week of disease spread, in order to reduce the global impact of the epidemic while keeping as many airports open as possible. 
This optimisation approach, based on a genetic algorithm, improved upon all of the other methods, hence providing a new model-based perspective on the decision making process that leads to the travel restrictions. 
One very interesting aspect of our results is that the algorithm leverages the complex structure of the network to place strategic ``fire-breaks'', which drastically reduce peak infections and total cases. 

\newpage

\appendix
\section{Appendix}
\subsection{Algorithm Derivation} \label{Algorithm_Derivation}
In this section we break down the various components of our diffusion and SEIRS models to describe in detail the mechanics of the epidemic simulator.
The pseudocode for the full procedure is provided in Table \ref{tab:Psuedocode}.

\paragraph{\textbf{SEIRS model.}}
The algorithm has been vectorised so that $\theta$ is a $4\times N$ matrix containing the states for all airports. 
The rows of $\theta$ are the $N$ dimensional vectors $S$, $E$, $I$ and $R$ representing the SEIRS compartments (for each of the airports' locations) with non-negative real numbers.

The fundamental equation for the local SEIRS is given by:
\begin{equation*}
 \begin{split}
\frac{dS}{dt} &= \delta R - \frac{S\beta I}{M} \\
\frac{dE}{dt} &= \frac{S\beta I}{M} - \epsilon E \\
\frac{dI}{dt} &= \epsilon E - \gamma I \\
\frac{dR}{dt} &= \gamma I - \delta R \\
 \end{split}
\end{equation*}

where $\beta$, $\epsilon$, $\gamma$ and $\delta$ are the epidemic parameters.
In the pseudocode, we make use of matrix notation and thus work with
$$
	\Sigma = \begin{pmatrix}
	-\beta & 0 & 0 & \delta \\
	\beta & -\epsilon & 0 & 0 \\
	0 & \epsilon & -\gamma & 0  \\
	0 & 0 & \gamma & -\delta  \\
	\end{pmatrix}
$$
With this notation, one step forward for the local SEIRS epidemic can be simply obtained with the matrix multiplication $\Sigma\theta$, for a suitable distribution matrix $\theta$.

\paragraph{\textbf{Diffusion model.}}
The mechanism for the diffusion model requires more steps and transformations which we describe in details here below. 
One main reference for the techniques used here is \cite{mark:10}.

If consider our network as a dynamic system, we have $N$ nodes, each with an associated amount of fluid $\psi_j \geq 0$.
The fluid is transferred from one node to another along the weighted edges of the network. The fluid flows from node $i$ to node $j$ at a rate proportional to the difference in the amount of fluid at each node $c(\psi_i - \psi_j)$, where $c$ is the constant of proportionality or more commonly referred to as the \textit{diffusion constant}.
This can be translated into matrix notation with:
\begin{equation}\label{app:eq:diff_1}
 \frac{d\psi_j}{dt} = c\sum_{i=1}^N (\psi_i - \psi_j)A_{ij}
\end{equation}
where $A$ is a suitable adjacency matrix. 
Equation \ref{app:eq:diff_1} can be rewritten as follows:
\begin{equation}
 \begin{split}
 \frac{d\psi_j}{dt} &= c\sum_{i=1}^N \psi_i A_{ij} - c\psi_j \text{ deg}(v_j)\\
  &= c(A - D)\psi\\
  &= -c(D - A)\psi
 \end{split}
\end{equation}
where $\text{ deg}(v_j)$ is the degree of node $j$, and $D$ is the $N\times N$ diagonal matrix with degrees as entries.

In our context, the fluid $\psi$ represents the local population available for travel, or, more precisely, a the size of a SEIRS compartment which is available for travel.
In fact, we have $4$ contemporaneous ``fluids'' that are moving through the system, corresponding to individuals for each of the compartments travelling between locations.
In order to accommodate this, we define a system through $\psi \in \mathbb{R}^{4\text{x}N}$ and redefine the equation as:
\begin{equation}\label{app:eq:diff_2}
 \frac{d\psi}{dt} =  -c\psi (D - A^\top) \in \mathbb{R}^{4\text{x}N}
\end{equation}
This completes a first step into defining the diffusion equation provided in the algorithm's pseudocode.

Equation \ref{app:eq:diff_2} simply defines how the local populations that are available for travelling \textit{should} travel, by defining the changes in size for each of the SEIRS compartments, at each of the locations.
However, the equation cannot be used as is, since it would create unrealistic travelling patterns.
The limitations that we address are the following:
\begin{enumerate}
 \item The migration process will have a stationary distribution, which likely does not correspond to the initial state of the system. This means that the populations that we observe at each locations will change dramatically over time, which is unrealistic. By contrast, we would like to have constant local populations $M_j$, which would be in line with our assumption of ``no permanent migration''.
 \item The total out-flow from a node may exceed the node's availability. Although unrealistic, the model allows for the number of outbound passengers to exceed the number of individuals that can travel from a location, so we need to add a condition to ensure that this does not happen.
 \item The number of available travellers should not be the only driving factor behind migration flows. As an example, Ethiopia has a very large population, which corresponds to a large number of potential travellers, however few routes go through this country. 
\end{enumerate}

As regards the first problem, we separate $\theta$ into $\theta_B$ and $\theta_T$, representing the \textbf{locals} and \textbf{visitors}, respectively.
This allows us to disentangle outbound and return travels, and monitor both. 
As initial condition, all local populations are in $\theta_B$, i.e. there are no visitors.
We proceed with the first step of the diffusion process by applying \eqref{app:eq:diff_2} on the matrix
$\psi_{+} = \theta_B \alpha$
which gives us the changes of local populations denoted by $\frac{d\psi_{+}}{dt}$. 
For those local populations that decrease due to \textit{outbound} individuals, we can write:
$\theta_B = \theta_B + \min (\frac{d\psi_{+}}{dt}, 0)$.
As regards the positive entries of $\frac{d\psi_{+}}{dt}$, which correspond to those locations that are supposed to see new individuals \textit{arriving}, we do not add these entries to $\theta_B$ but to $\theta_T$.
This is in agreement with the fact that these travellers will join the populations of visitors in their new locations, and not the locals.
In mathematical terms:
$\theta_T = \theta_T + \max (\frac{d\psi_{+}}{dt}, 0)$.

Analogously, we apply the diffusion process \eqref{app:eq:diff_2} to the matrix $\psi_{-} = \theta_T \alpha$, which gives us the changes of visitor populations denoted by $\frac{d\psi_{-}}{dt}$.
Then, the negative entries of this term will correspond to the visitors that are \textit{departing}, whereas the positive entries are the visitors that are returning at the end of their travels, thus flowing back into the local populations.
The complete updates are then:
\begin{equation*}
 \begin{split}
  \theta_B &= \theta_B + \min \left( \frac{d\psi_{+}}{dt}, 0 \right) + \max \left( \frac{d\psi_{-}}{dt}, 0 \right) \\
  \theta_T &= \theta_T + \max \left( \frac{d\psi_{+}}{dt}, 0 \right) + \min \left( \frac{d\psi_{-}}{dt}, 0 \right)
 \end{split}
\end{equation*}
Note that we have carefully chosen the wording \textit{outbound},  \textit{arriving},  \textit{departing},  \textit{returning} to reflect the sequence of steps all travellers pass through in order. This is crucial to ensure logical behaviour of travellers and also to prevent leakage of fluid / people from the network.\\

In order to address the second issue we can simply divide $\psi_i$ by the degree of the node which ensures that the outflows computed to every other node will be at most $\psi_i$. 
This results in the diffusion equation \eqref{app:eq:diff_2} being changed to 
\begin{equation}\label{app:eq:diff_3}
 \frac{d\psi}{dt} =  -c\psi D^{-1}(D - A^\top) \in \mathbb{R}^{4\text{x}N}
\end{equation}
This arbitrary operation simply guarantees a rescaling of each individual value, so that we can ensure that the entries of $\psi$ remain non-negative throughout the procedure.
\\

As regards the third issue, this cannot be addressed with a simple rescaling as we need to ensure that the importance of airport is properly captured by our approach. 
For this reason, we replace the adjacency matrix $A$ with a weighted matrix $C$ to encourage flows to more central airports.
We construct $C$ by first amplifying the entries of $A$ by the centrality of the receiving node, and then we rescale the entries to obtain a row stochastic matrix.
The generic element of $C$ is defined as follows:
$$C_{ij} = \frac{A_{ij}P_j}{\sum_{k=1}^N A_{ik}P_k}$$
where $P_j$ is the page-rank centrality of node $v_j$.
This makes the migration matrix more realistic since it captures better the role played by each airport, while maintaining exactly the same adjacency structure as informed by $A$.
We then simply make the matrix replacement in \eqref{app:eq:diff_2}, noting that, since $C$ is row stochastic, the matrix $D$ becomes the identity matrix, and thus disappears from the equation.
Hence, the diffusion \eqref{app:eq:diff_2} is replaced by:
$$\frac{d\psi}{dt} =  -c\psi (I - C^\top) \in \mathbb{R}^{4\text{x}N}$$
which is equal to the formula provided in the pseudocode.


\printbibliography

%
%

\end{document}